\documentclass[envcountsame,envcountsect]{llncs}
\usepackage[letterpaper,hmargin=1.3in,vmargin=1.3in]{geometry}

\usepackage{graphicx}
\usepackage{amssymb}
\usepackage{epstopdf}
\usepackage{amsmath}
\usepackage{qip}
\usepackage{framed}
\usepackage{epsfig}

\newcommand{\ot}{\ensuremath{\mbox{{\sc {\small 1-2}-ot}}}}

\newcommand{\otalone}{{\sc ot}}
\newcommand{\nlb}{{\sc nl}}

\newcommand{\emb}[1]{\ensuremath{\mathcal{E}}(#1)}

\newcommand{\expe}[1]{\ensuremath{\mbox{e}^{#1}}}
\newcommand{\dep}[2]{\ensuremath{#1 \searrow #2}}
\newcommand{\otp}[1]{\ensuremath{\ot^{#1}}}
\newcommand{\srotp}[1]{\ensuremath{\mbox{\sc rot}^{#1}}}
\newcommand{\srot}{\srotp{r}}
\newcommand{\sotr}{\otp{r}}

\newcommand{\sand}{\ensuremath{\mbox{\sc sand}}}

\newcommand{\otn}{\ensuremath{\ot_p}}

\newcommand{\pnl}{\ensuremath{P^{\mbox{{\tiny \nlb}}}_{X,Y}}}
\newcommand{\pot}{\ensuremath{P^{\mbox{{\tiny \otalone}}}_{X,Y}}}

\newcommand{\psot}{\ensuremath{P^{\mbox{{\tiny {\sc ot}$^r$}}}_{X,Y}}}
\newcommand{\potn}{\ensuremath{P^{\mbox{{\tiny \otalone$_p$}}}_{X,Y}}}
\newcommand{\potnq}{\ensuremath{P^{\mbox{{\tiny \otalone$_{1/4}$}}}_{X,Y}}}

\newcommand{\psrot}{\ensuremath{P^{\mbox{{\tiny {\sc rot}$^r$}}}_{X,Y}}}

\newcommand{\hil}{\mathcal{H}}

\newcommand{\append}[2]{#1}

\newcommand{\myparagraph}[1]{\medskip\noindent{\sc #1}}

\newcommand{\assign}{\ensuremath{\kern.5ex\raisebox{.1ex}{\mbox{\rm:}}\kern -.3em =}}

\newcounter{itm}

\title{
On the Power of\\Two-Party Quantum Cryptography
}

\author{
Louis Salvail\inst{1}\fnmsep\thanks{supported by QUSEP (funded by the 
                                    Danish Natural Science Research Council), Canada's NSERC, and the QuantumWorks network.} 
\and
Christian Schaffner\inst{2}\fnmsep\thanks{supported by EU fifth framework project QAP IST 015848 and the NWO VICI project 2004-2009} 
 \and Miroslava Sot\'{a}kov\'{a}\inst{3}
}

\institute{
Universit\'e de Montr\'eal (DIRO), QC, Canada\\
\email{salvail@iro.umontreal.ca}
\and
Centrum Wiskunde \& Informatica (CWI) Amsterdam, The Netherlands\\
\email{c.schaffner@cwi.nl}
\and
SUNY Stony Brook (Dept. of Computer Science), NY, USA\\
\email{mirka@cs.au.dk}
}

\begin{document}
\pagestyle{plain}

\maketitle

\begin{abstract}
  We study quantum protocols among two distrustful parties. Under the
  sole assumption of correctness---guaranteeing that honest players
  obtain their correct outcomes---we show that every protocol
  implementing a non-trivial primitive necessarily leaks information
  to a dishonest player. This extends known impossibility results to
  all non-trivial primitives. We provide a framework for quantifying
  this leakage and argue that leakage is a good measure for the
  privacy provided to the players by a given protocol. Our framework
  also covers the case where the two players are helped by a trusted
  third party. We show that despite the help of a trusted third party,
  the players cannot amplify the cryptographic power of any
  primitive. All our results hold even against quantum
  honest-but-curious adversaries who honestly follow the protocol but
  purify their actions and apply a different measurement at the end of
  the protocol. As concrete examples, we establish lower bounds on the
  leakage of standard universal two-party primitives such as oblivious
  transfer.

\vspace{1mm}

\end{abstract}

\section{Introduction}
\label{chap:intro}
Quantum communication allows to implement tasks which are classically
impossible. The most prominent example is quantum key
distribution~\cite{BB84} where two honest players establish a secure
key against an eavesdropper. In the two-party setting however, quantum
and classical cryptography often show similar limits. Oblivious
transfer~\cite{Lo97}, bit commitment~\cite{Mayers97,LC97}, and even fair
coin tossing~\cite{Kitaev03} are impossible to realize securely both
classically and quantumly.  On the other hand, quantum cryptography
allows for some weaker primitives impossible in the classical
world. For example, quantum coin-flipping protocols with maximum bias
of $\frac{1}{\sqrt{2}}-\frac12$ exist\footnote{In fact, protocols with
  better bias are known for weak quantum coin
  flipping~\cite{Mochon04,Mochon05,Mochon07}.} against any
adversary~\cite{CK09} while remaining impossible based solely on
classical communication. A few other weak primitives are known to be
possible with quantum communication.
For example, the generation of an additive
secret-sharing for the product $xy$ of two bits, where Alice holds bit
$x$ and Bob bit $y$, has been introduced by Popescu and Rohrlich as
machines modeling non-signaling non-locality (also called NL-boxes)~\cite{PR94}.
If Alice and Bob share an EPR pair, they can simulate an NL-box with symmetric
error probability $\sin^{2}{\frac{\pi}{8}}$~\cite{PR94,BLMPPR05}.
Equivalently, Alice and Bob can implement {\em 1-out-of-2 oblivious
  transfer} (\ot) privately provided the receiver Bob gets the bit of
his choice only with probability of error $\sin^{2}{\frac{\pi}{8}}$~\cite{Ambainis05OT}.  It is easy to verify that even with such imperfection these two
primitives are impossible to realize in the classical world.
This discussion naturally leads to the following question:
\begin{itemize}
\item Which two-party cryptographic primitives are possible to achieve using
quantum communication?  
\end{itemize}
Most standard classical two-party primitives have been shown
impossible to implement securely against weak quantum adversaries
reminiscent to the classical honest-but-curious (HBC)
behavior~\cite{Lo97}. The idea behind these impossibility proofs is to
consider parties that {\em purify} their actions throughout the
protocol execution.  This behavior is indistinguishable from the one
specified by the protocol but guarantees that the joint quantum state
held by Alice and Bob at any point during the protocol remains
pure. The possibility for players to behave that way in any two-party
protocol has important consequences. For instance, the impossibility
of quantum bit commitment follows from this fact~\cite{Mayers97,LC97}:
After the commit phase, Alice and Bob share the pure state
$\ket{\psi^x} \in \hil_A \otimes \hil_B$ corresponding to the
commitment of bit~$x$. Since a proper commitment scheme provides no
information about $x$ to the receiver Bob, it follows that
$\tr_A\proj{\psi^{{0}}} =\tr_A\proj{\psi^1}$. In this case, the
Schmidt decomposition guarantees that there exists a unitary $U_{0,1}$
acting only on Alice's side such that $\ket{\psi^1} = (U_{0,1}\otimes
\I_{B})\ket{\psi^{{0}}}$.  In other words, if the commitment is
concealing then Alice can open the bit of her choice by applying a
suitable unitary transform only to her part. A similar argument allows
to conclude that \ot\ is impossible~\cite{Lo97}: Suppose Alice is
sending the pair of bits $(b_0,b_1)$ to Bob through \ot.  Since Alice
does not learn Bob's selection bit, it follows that Bob can get bit
$b_0$ before undoing the reception of $b_0$ and transforming it into
the reception of $b_1$ using a local unitary transform similar to
$U_{0,1}$ for bit commitment. For both these primitives, privacy for
one player implies that local actions by the other player can
transform the honest execution with one input into the honest
execution with another input.

In this paper, we investigate the cryptographic power of two-party
quantum protocols against players that purify their actions. This {\em
  quantum honest-but-curious (QHBC)} behavior is the natural quantum
version of classical HBC behavior.  We consider the setting where
Alice obtains random variable $X$ and Bob random variable $Y$
according to the joint probability distribution $P_{X,Y}$. Any
$P_{X,Y}$ models a two-party cryptographic primitive where neither
Alice nor Bob provide input. For the purpose of this paper, this model
is general enough since any two-party primitive with inputs can be
randomized (Alice and Bob pick their input at random) so that its
behavior can be described by a suitable joint probability distribution
$P_{X,Y}$. If the randomized version $P_{X,Y}$ is shown to be
impossible to implement securely by any quantum protocol then
also the original primitive with inputs is impossible.

Any quantum protocol implementing $P_{X,Y}$ must produce, when both
parties purify their actions, a joint pure state $\ket{\psi}\in
\hil_{AA'} \otimes\hil_{BB'}$ that, when subsystems of $A$ and $B$ are
measured in the computational basis, leads to outcomes $X$ and $Y$
according the distribution $P_{X,Y}$. Notice that the registers $A'$
and $B'$ only provide the players with extra working space and, as
such, do not contribute to the output of the functionality (so parties
are free to measure them the way they want). In this paper, we adopt a
somewhat strict point of view and define a quantum protocol $\pi$ for
$P_{X,Y}$ to be \emph{correct} if and only if the correct outcomes
$X,Y$ are obtained \emph{and} the registers $A'$ and $B'$ do not
provide any additional information about $Y$ and $X$ respectively
since otherwise $\pi$ would be implementing a different primitive
$P_{XX',YY'}$ rather than $P_{X,Y}$. 

The state $\ket{\psi}$ produced by any correct protocol for $P_{X,Y}$
is called a {\em quantum embedding} of $P_{X,Y}$.  An embedding is
called \emph{regular} if the registers $A'$ and $B'$ are empty. Any
embedding $\ket{\psi}\in \hil_{AA'}\otimes\hil_{BB'}$ can be produced
in the QHBC model by the trivial protocol asking Alice to generate
$\ket{\psi}$ before sending the quantum state in $\hil_{BB'}$ to
Bob. Therefore, it is sufficient to investigate the cryptographic
power of embeddings in order to understand the power of two-party
quantum cryptography in the QHBC model.

Notice that if $X$ and $Y$ were provided privately to Alice and
Bob---through a trusted third party for instance---then the expected
amount of information one party gets about the other party's output is
minimal and can be quantified by the Shannon mutual information
$I(X;Y)$ between $X$ and $Y$. Assume that $\ket{\psi}\in
\hil_{AA'}\otimes\hil_{BB'}$ is the embedding of $P_{X,Y}$ produced by
a correct quantum protocol.  We define the leakage of $\ket{\psi}$ as
\begin{equation}
\label{leak_intro} 
\Delta_{\psi} \assign \max \left\{ \, S(X;BB')-I(X;Y) \, ,
  \:S(Y;AA')-I(Y;X) \,\right\},
\end{equation}
where $S(X;BB')$ (resp. $S(Y;AA')$)  is the information
the quantum registers $BB'$ (resp. $AA'$) provide about 
the output $X$ (resp. $Y$). That is, the leakage is the
maximum amount of extra information about the other party's output
given the quantum state held by one party. It turns out that
$S(X;BB')=S(Y;AA')$ holds for all embeddings, exhibiting a symmetry
similar to its classical counterpart $I(X;Y)=I(Y;X)$ and therefore,
the two quantities we are taking the maximum of (in the definition of
leakage above) coincide. 

\myparagraph{Contributions. } Our first contribution establishes that
the notion of leakage is well behaved.  We show that the leakage of
any embedding for $P_{X,Y}$ is lower bounded by the leakage of some
regular embedding of the same primitive. Thus, in order to lower bound
the leakage of any correct implementation of a given primitive, it
suffices to minimize the leakage over all its regular embeddings.  We
also show that the only non-leaking embeddings are the ones for
trivial primitives, where a primitive $P_{X,Y}$ is said to be {\em
  (cryptographically) trivial} if it can be generated by a classical
protocol against HBC adversaries\footnote{\label{foot:caveat}We are
  aware of the fact that our definition of triviality encompasses
  cryptographically interesting primitives like coin-tossing and
  generalizations thereof for which highly non-trivial protocols
  exist~\cite{Mochon07,CK09}. However, the important fact (for
  the purpose of this paper) is that all these primitives can be
  implemented by \emph{trivial} classical protocols against HBC
  adversaries.}. It follows that any quantum protocol implementing a
non-trivial primitive $P_{X,Y}$ must leak information under the sole
assumption that it produces $(X,Y)$ with the right joint
distribution. This extends known impossibility results for two-party
primitives to all non-trivial primitives.

Embeddings of primitives arise from protocols where Alice and Bob have
full control over the environment. Having in mind that any embedding
of a non-trivial primitive leaks information, it is natural to
investigate what tasks can be implemented without leakage with the
help of a trusted third party. The notion of leakage can easily be
adapted to this scenario.
We show that no cryptographic two-party primitive can be implemented
without leakage with just one call to the ideal functionality of a
weaker primitive\footnote{The weakness of a primitive will be formally
  defined in terms of entropic monotones for classical two-party
  computation introduced by Wolf and Wullschleger~\cite{WW04}, see
  Section~\ref{sec:limitedresources}.}. This new impossibility result
does not follow from the ones known since they all assume that the
state shared between Alice and Bob is pure.

We then turn our attention to the leakage of correct protocols for a
few concrete universal primitives. From the results described above,
the leakage of any correct implementation of a primitive can be
determined by finding the (regular) embedding that minimizes the
leakage.  In general, this is not an easy task since it requires to
find the eigenvalues of the reduced density matrix $\rho_A =
\tr_B\proj{\psi}$ (or equivalently $\rho_B = \tr_A\proj{\psi}$). As
far as we know, no known results allow us to obtain a non-trivial
lower bound on the leakage (which is the difference between the mutual
information and accessible information) of non-trivial primitives. One
reason being that in our setting we need to lower bound this
difference with respect to a measurement in one particular basis.
However, when $P_{X,Y}$ is such that the bit-length of either $X$ or
$Y$ is short, the leakage can be computed precisely. We show that any
correct implementation of \ot\ necessarily leaks $\frac{1}{2}$\,bit.
Since NL-boxes and \ot\ are locally equivalent, the same minimal
leakage applies to NL-boxes~\cite{WW05b}.  This is a stronger
impossibility result than the one by Lo~\cite{Lo97} since he
assumes perfect/statistical privacy against one party while our
approach only assumes correctness (while both approaches apply even
against QHBC adversaries). We finally show that for Rabin-OT and \ot\
of $r$-bit strings (i.e. \srot\ and \sotr\ respectively), the leakage
approaches $1$ exponentially in $r$. In other words, correct
implementations of these two primitives trivialize as $r$ increases
since the sender gets almost all information about Bob's reception of
the string (in case of \srot) and Bob's choice bit (in case of
\sotr). These are the first quantitative impossibility results for
these primitives and certainly the first time the hardness of
implementing different flavors of string OTs is shown to increase as
the strings to be transmitted get longer.

Finally, we note that our lower bounds on the leakage of the
randomized primitives also lower-bound the minimum leakage for the
standard versions of these primitives\footnote{The definition of
  leakage of an embedding can be generalized to protocols with inputs,
  where it is defined as $\max\{ \sup_{V_B} S(X;V_B)-I(X;Y) \, , \,
  \sup_{V_A} S(V_A;Y)-I(X;Y) \}$, where $X$ and $Y$ involve both 
inputs and outputs of Alice and Bob, respectively. The supremum is taken over all
  possible (quantum) views $V_A$ and $V_B$ of Alice and Bob obtained
  by their (QHBC-consistent) actions (and containing their inputs).}
where the players choose their
inputs uniformly at random. While we focus on the typical case where
the primitives are run with uniform inputs, the same reasoning can be
applied to primitives with arbitrary distributions of inputs.

\myparagraph{Related Work. } Our framework allows to quantify the
minimum amount of leakage whereas standard impossibility proofs as the
ones of~\cite{LC97,Mayers97,Lo97,AKSW07,BCS09prep} do not in general
provide such quantification since they usually assume privacy for one
player in order to show that the protocol must be totally insecure for
the other player\footnote{Trade-offs between the security for one and the
  security for the other player have been considered before, but
  either the relaxation of security has to be very small~\cite{Lo97}
  or the trade-offs are restricted to particular primitives such as
  commitments~\cite{SR01,BCHLW08}.}. By contrast, we derive lower
bounds for the leakage of any correct implementation.  At first
glance, our approach seems contradictory with standard impossibility
proofs since embeddings leak the same amount towards both parties.  To
resolve this apparent paradox it suffices to observe that in previous
approaches only the adversary purified its actions whereas in our case
both parties do. If a honest player does not purify his actions then
some leakage may be lost by the act of irreversibly and unnecessarily
measuring some of his quantum registers.

Our results complement the ones obtained by Colbeck
in~\cite{Colbeck07} for the setting where Alice and Bob have inputs
and obtain identical outcomes (called single-function computations).
\cite{Colbeck07} shows that in any correct implementation of
primitives of a certain form, an honest-but-curious player can access
more information about the other party's input than it is available
through the ideal functionality. Unlike~\cite{Colbeck07}, we deal in
our work with the case where Alice and Bob do not have inputs but
might receive different outputs according to a joint probability
distributions. We show that only trivial distributions can be
implemented securely in the QHBC model.  Furthermore, we introduce a
quantitative measure of protocol-insecurity that lets us answer which
embedding allow the least effective cheating.

Another notion of privacy in quantum protocols, generalizing its
classical counterpart from~\cite{BK91,Kushilevitz92}, is proposed by
Klauck in~\cite{Klauck04}.  Therein, two-party quantum protocols with
inputs for computing a function
$f:\mathcal{X}\times\mathcal{Y}\rightarrow \mathcal{Z}$, where
$\mathcal{X}$ and $\mathcal{Y}$ denote Alice's and Bob's respective
input spaces, and privacy against QHBC adversaries are considered.
Privacy of a protocol is measured in terms of \emph{privacy loss},
defined for each round of the protocol and fixed distribution of
inputs $P_{X',Y'}$ by $S(B;X|Y)=H(X|Y)-S(X|B,Y)$, where $B$ denotes
Bob's private working register, and $X\assign(X',f(X',Y'))$,
$Y\assign(Y',f(X',Y'))$ represent the complete views of Alice and Bob,
respectively.  Privacy loss of the entire protocol is then defined as
the supremum over all joint input distributions, protocol rounds, and
states of working registers.  In our framework, privacy loss
corresponds to $S(X;YB)-I(X;Y)$ from Alice point's of view and
$S(Y;XA)-I(X;Y)$ from Bob's point of view.  Privacy loss is therefore
very similar to our definition of leakage except that it requires the
players to get their respective honest outputs.  As a consequence, the
protocol implementing $P_{X,Y}$ by asking one party to prepare a
regular embedding of $P_{X,Y}$ before sending her register to the
other party would have no privacy loss. Moreover, the scenario
analyzed in \cite{Klauck04} is restricted to primitives which provide
the same output $f(X,Y)$ to both players.  Another difference is that
since privacy loss is computed over all rounds of a protocol, a party
is allowed to abort which is not considered QHBC in our setting.  In
conclusion, the model of~\cite{Klauck04} is different from ours even
though the measures of privacy loss and leakage are similar.
\cite{Klauck04} provides interesting results concerning trade-offs
between privacy loss and communication complexity of quantum
protocols, building upon similar results of~\cite{BK91,Kushilevitz92}
in the classical scenario. It would be interesting to know whether a
similar operational meaning can also be assigned to the new measure of
privacy, introduced in this paper.  

A recent result by K\"unzler et al.~\cite{KMR09} shows that two-party
functions that are securely computable against active quantum
adversaries form a strict subset of the set of functions which are
securely computable in the classical HBC model. This complements our
result that the sets of securely computable functions in both HBC and
QHBC models are the same.

\myparagraph{Roadmap.}  In Section~\ref{chap:prelim}, we introduce the
cryptographic and information-theoretic notions and concepts used
throughout the paper. We define, motivate, and analyze the generality
of modeling two-party quantum protocols by embeddings in
Section~\ref{qembprotocols} and define triviality of primitives and
embeddings. In Section~\ref{crypto}, we define the notion of leakage
of embeddings, show basic properties and argue that it is a reasonable
measure of privacy. In Section~\ref{sec:primleakage}, we explicitly
lower bound the leakage of some universal two-party
primitives. Finally, in Section~\ref{conclusion} we discuss possible
directions for future research and open questions.

\section{Preliminaries}
\label{chap:prelim}
\myparagraph{Quantum Information Theory.}  Let $\ket{\psi}_{AB} \in
\hil_{AB}$ be an arbitrary pure state of the joint systems $A$ and
$B$. The states of these subsystems are $\rho_A = \tr_B\proj{\psi}$
and $\rho_B=\tr_A\proj{\psi}$, respectively.  We denote by $S(A)
\assign S(\rho_A)$ and $S(B) \assign S(\rho_B)$ the von Neumann
entropy (defined as the Shannon entropy of the eigenvalues of the
density matrix) of subsystem $A$ and $B$ respectively. Since the joint
system is in a pure state, it follows from the Schmidt decomposition
that $S(A)=S(B)$ (see e.g.~\cite{NC00}). Analogously to their
classical counterparts, we can define quantum conditional entropy
$S(A|B)\assign S(AB)-S(B)$, and quantum mutual information
$S(A;B)\assign S(A)+S(B)-S(AB)=S(A)-S(A|B).$ Even though in general,
$S(A|B)$ can be negative, $S(A|B)\geq 0$ is always true if $A$ is a
classical register.  Let $R=\{(P_X(x),\rho_R^x\}_{x\in {\cal X}}$ be
an ensemble of states $\rho_R^x$ with prior probability $P_X(x)$. The
average quantum state is $\rho_R = \sum_{x\in {\cal X}} P_X(x)
\rho_R^x$.  The famous result by Holevo upper-bounds the amount of
classical information about $X$ that can be obtained by measuring
$\rho_R$:
\begin{theorem}[Holevo bound~\cite{Holevo73,Ruskai02}]\label{holevo}
  Let $Y$ be the random variable describing the outcome of some
  measurement applied to $\rho_R$ for $R=\{P_X(x),\rho_R^x\}_{x\in
    {\cal X}}$. Then, $ I(X;Y) \leq S(\rho_R)-\sum_xP_X(x)S(\rho_R^x),$
  where equality can be achieved if and only if
  $\{\rho_R^x\}_{x\in{\cal X}}$ are simultaneously diagonalizable.
\end{theorem}
Note that if all states in the ensemble are pure and all different
then in order to achieve equality in the theorem above, they have to
form an orthonormal basis of the space they span. In this case, the
variable $Y$ achieving equality is the measurement outcome in this
orthonormal basis.

\myparagraph{Dependent Part.}  The following definition introduces a
random variable describing the correlation between two random
variables $X$ and $Y$, obtained by collapsing all values $x_1$ and
$x_2$ for which $Y$ has the same conditional distribution, to a
single value.
\begin{definition}[Dependent part~\cite{WW04}]
\label{dep_part}
For two random variables $X,Y$, let $f_X(x) \assign P_{Y|X=x}$. Then the
\emph{dependent part of $X$ with respect to $Y$} is defined as $\dep{X}{Y} \assign f_X(X)$.
\end{definition}
The dependent part $\dep{X}{Y}$ is the minimum random variable among
the random variables computable from $X$ for which $X\leftrightarrow
\dep{X}{Y} \leftrightarrow Y$ forms a Markov chain \cite{WW04}. In
other words, for any random variable $K=f(X)$ such that
$X\leftrightarrow K \leftrightarrow Y$ is a Markov chain, there exists
a function $g$ such that $g(K)=\dep{X}{Y}$.  Immediately from the
definition we get several other properties of
$\dep{X}{Y}$~\cite{WW04}: $H(Y|\dep{X}{Y})=H(Y|X)$,
$I(X;Y)=I(\dep{X}{Y};Y)$, and $\dep{X}{Y}=\dep{X}{(\dep{Y}{X})}$. The
second and the third formula yield $I(X;Y)=I(\dep{X}{Y};\dep{Y}{X})$.

The notion of dependent part has been further investigated in
\cite{FWW04,IMNW04,WW05a}.  Wullschleger and Wolf have shown that
quantities $H(X\searrow Y|Y)$ and $H(Y\searrow X|X)$ are monotones for
two-party computation~\cite{WW05a}. That is, none of these values can
increase during classical two-party protocols. In particular, if Alice
and Bob start a protocol from scratch then classical two-party
protocols can only produce $(X,Y)$ such that: $H(X\searrow
Y|Y)=H(Y\searrow X|X)=0$, since $H(X\searrow Y|Y)>0$ if and only if
$H(Y\searrow X|X)>0$~\cite{WW05a}.  Conversely, any primitive
satisfying $H(X\searrow Y|Y)=H(Y\searrow X|X)=0$ can be implemented
securely in the honest-but-curious (HBC) model. We call such
primitives \emph{trivial}\footnote{See Footnote~\ref{foot:caveat} for
  a caveat about this terminology.}.

\myparagraph{Purification.}
\label{purification}
All security questions we ask are with respect to \emph{(quantum)
  honest-but-curious} adversaries. In the classical honest-but-curious
adversary model (HBC), the parties follow the instructions of a
protocol but store all information available to them. Quantum
honest-but-curious adversaries (QHBC), on the other hand, are allowed
to behave in an arbitrary way that cannot be distinguished from their
honest behavior by the other player.

Almost all impossibility results in quantum cryptography rely upon a
quantum honest-but-curious behavior of the adversary.  This behavior
consists in {\em purifying} all actions of the honest
players. Purifying means that instead of invoking classical randomness
from a random tape, for instance, the adversary relies upon quantum
registers holding all random bits needed. The operations to be
executed from the random outcome are then performed quantumly without
fixing the random outcomes.  For example, suppose a protocol instructs
a party to pick with probability $p$ state $\ket{\phi^0}_C$ and with
probability $1-p$ state $\ket{\phi^1}_C$ before sending it to the
other party through the quantum channel $C$. The purified version of
this instruction looks as follows: Prepare a quantum register in state
$\sqrt{p}\ket{0}_R+\sqrt{1-p}\ket{1}_R$ holding the random
process. Add a new register initially in state $\ket{0}_C$ before
applying the unitary transform $U:\ket{r}_R\ket{0}_C \mapsto
\ket{r}_R\ket{\phi^r}_C$ for $r\in\{0,1\}$, send register $C$
through the quantum channel and keep register $R$.

From the receiver's point of view, the purified behavior is
indistinguishable from the one relying upon a classical source of
randomness because in both cases, the state of register $C$ is
$\rho=p\proj{\phi^0}+(1-p)\proj{\phi^1}$. All operations invoking
classical randomness can be purified
similarly~\cite{LC97,Mayers97,Lo97,Kent04}. The result is that
measurements are postponed as much as possible and only extract
information required to run the protocol in the sense that only when
both players need to know a random outcome, the corresponding quantum
register holding the random coin will be measured.  If both players
purify their actions then the joint state at any point during the
execution will remain pure, until the very last step of the protocol
when the outcomes are measured.

\myparagraph{Secure Two-Party Computation.}
\label{prel.crypto}
In Section~\ref{sec:primleakage}, we investigate the leakage of
several universal cryptographic two-party primitives. By universality
we mean that any two-party secure function evaluation can be reduced
to them. We investigate the completely randomized versions where
players do not have inputs but receive randomized outputs instead.
Throughout this paper, the term \emph{primitive} usually refers to the
joint probability distribution defining its randomized version. Any
protocol implementing the standard version of a primitive (with
inputs) can also be used to implement a randomized version of the same
primitive, with the ``inputs'' chosen according to an arbitrary fixed
probability distribution.

\section{Two-Party Protocols and Their Embeddings}\label{qembprotocols}

\subsection{Correctness}\label{protocols}
In this work, we consider \emph{cryptographic primitives} providing
$X$ to honest player Alice and $Y$ to honest player Bob according to a
joint probability distribution $P_{X,Y}$. The goal of this section is
to define when a protocol $\pi$ \emph{correctly implements} the
primitive $P_{X,Y}$. The first natural requirement is that once the
actions of $\pi$ are purified by both players, measurements of
registers $A$ and $B$ in the computational basis\footnote{It is clear
  that every quantum protocol for which the final measurement
  (providing $(x,y)$ with distribution $P_{X,Y}$ to the players) is
  not in the computational basis can be transformed into a protocol of
  the described form by two additional local unitary transformations.}
provide joint outcome $(X,Y)=(x,y)$ with probability $P_{X,Y}(x,y)$.

Protocol $\pi$ can use extra registers $A'$ on Alice's and $B'$ on
Bob's side providing them with (quantum) working space.  The
purification of all actions of $\pi$ therefore generates a pure state
$\ket{\psi}\in \hil_{AB}\otimes \hil_{A'B'}$. A second requirement for
the correctness of the protocol $\pi$ is that these extra registers
are only used as working space, i.e.~the final state
$\ket{\psi}_{ABA'B'}$ is such that the content of Alice's working
register $A'$ does not give her any further information about Bob's
output $Y$ than what she can infer from her honest output $X$ and vice
versa for $B'$. Formally, we require that $S(XA' ; Y)=I(X;Y)$ and $S(X
; YB')=I(X;Y)$ or equivalently, that $A'\leftrightarrow X
\leftrightarrow Y$ and $X\leftrightarrow Y \leftrightarrow B'$ form
Markov chains\footnote{Markov chains with quantum ends have been
  defined in~\cite{DFSS07} and used in subsequent works such
  as~\cite{FS09}. It is straightforward to verify that the entropic
  condition $S(XA' ; Y)=I(X;Y)$ is equivalent to $A'\leftrightarrow X
  \leftrightarrow Y$ being a Markov chain and similarly for the other
  condition.}.
\begin{definition}\label{defcorrect}
\label{correct}
A protocol $\pi$ for $P_{X,Y}$ is \emph{correct} if measuring
registers $A$ and $B$ of its final state in the computational basis
yields outcomes $X$ and $Y$ with distribution $P_{X,Y}$ and the final
state satisfies $S(X; YB') = S(XA'; Y) =I(X;Y)$ where $A'$ and $B'$
denote the extra working registers of Alice and Bob. The state
$\ket{\psi}\in \hil_{AB} \otimes \hil_{A'B'}$ is called an
\emph{embedding of $P_{X,Y}$} if it can be produced by the
purification of a correct protocol for $P_{X,Y}$.
\end{definition}
We would like to point out that our definition of correctness is
stronger than the usual classical notion which only requires the
correct distribution of the output of the honest players. For example,
the trivial classical protocol for the primitive $P_{X,Y}$ in which
Alice samples both player's outputs $XY$, sends $Y$ to Bob, but keeps
a copy of $Y$ for herself, is \emph{not correct} according to our
definition, because it implements a fundamentally different primitive,
namely $P_{XY,Y}$.

\subsection{Regular Embeddings} \label{sec:embed} We call an embedding
$\ket{\psi}_{ABA'B'}$ \emph{regular} if the working registers $A',B'$
are empty. Formally, let ${\Theta}_{n,m} \assign
\{\theta:\{0,1\}^n\times\{0,1\}^m \rightarrow [0\ldots2\pi) \}$ be the
set of functions mapping bit-strings of length $m+n$ to real numbers
between $0$ and $2\pi$.

\begin{definition}
For a joint probability distribution $P_{X,Y}$ where $X\in\{0,1\}^n$
and $Y\in\{0,1\}^m$, we define the set
\[ \emb{P_{X,Y}} \assign \left\{\ket{\psi} \in \hil_{AB} : \ket{\psi}
  = \!\!\!\!\!\!\!\! \sum_{x\in\{0,1\}^n \! , \, y \in\{0,1\}^m}
  \!\!\!\!\!\!\!\!
  \expe{i\theta(x,y)}\sqrt{P_{X,Y}(x,y)}\ket{x,y}_{AB} \, , \theta\in
  {\Theta}_{n,m} \right\} \, ,
\]
and call any state $\ket{\psi}\in \emb{P_{X,Y}}$ a {\em regular
  embedding} of the joint probability distribution $P_{X,Y}$ .
\end{definition}

Clearly, any $\ket{\psi}\in \emb{P_{X,Y}}$ produces $(X,Y)$ with
distribution $P_{X,Y}$ since the probability that Alice measures $x$
and Bob measures $y$ in the computational basis is
$|\bracket{\psi}{x,y}|^2= P_{X,Y}(x,y)$. In order to specify a
particular regular embedding one only needs to give the description of
the {\em phase function} $\theta(x,y)$. We denote by
$\ket{\psi_\theta}\in \emb{P_{X,Y}}$ the quantum embedding of
$P_{X,Y}$ with phase function~$\theta$.  The constant function
$\theta(x,y) \assign 0$ for all $x\in\{0,1\}^n, y\in\{0,1\}^m$
corresponds to what we call {\em canonical embedding}
$\ket{\psi_{\vec{0}}} \assign
\sum_{x,y}\sqrt{P_{X,Y}(x,y)}\ket{x,y}_{AB}$ .

In Lemma~\ref{super_leak} below we show that every primitive $P_{X,Y}$
has a regular embedding which is in some sense the most secure among
all embeddings of $P_{X,Y}$.

\subsection{Trivial Classical Primitives and Trivial Embeddings}
In this section, we define \emph{triviality} of classical primitives
and (bipartite) embeddings. We show that for any non-trivial classical
primitive, its canonical quantum embedding is also non-trivial.
Intuitively, a primitive $P_{X,Y}$ is {\em trivial} if $X$ and $Y$ can
be generated by Alice and Bob from scratch in the classical
honest-but-curious (HBC) model\footnote{See Footnote~\ref{foot:caveat}
  for a caveat about this terminology.}. Formally, we define
triviality via an entropic quantity based on the notion of
\emph{dependent part} (see Section~\ref{chap:prelim}).
\begin{definition}
  A primitive $P_{X,Y}$ is called {\em trivial} if it satisfies
  $H(\dep{X}{Y}|Y)=0$, or equivalently,
  \mbox{$H(\dep{Y}{X}|X)=0$}. Otherwise, the primitive is called {\em
    non-trivial}.
\end{definition}

\begin{definition}
  A regular embedding $\ket{\psi}_{AB}\in \emb{P_{X,Y}}$ is called
  {\em trivial} if either $S(\dep{X}{Y}|B)=0$ or
  $S(\dep{Y}{X}|A)=0$. Otherwise, we say that $\ket{\psi}_{AB}$ is
  {\em non-trivial}.
\end{definition}
Notice that unlike in the classical case,
$S(\dep{X}{Y}|B)=0\Leftrightarrow S(\dep{Y}{X}|A)=0$ does not hold in
general. As an example, consider a shared quantum state where the
computational basis corresponds to the Schmidt basis for only one of
its subsystems, say for $A$. Let
$\ket{\psi}=\alpha\ket{0}_A\ket{\xi_0}_B+\beta\ket{1}_A\ket{\xi_1}_B$
be such that both subsystems are two-dimensional,
$\{\ket{\xi_0},\ket{\xi_1}\}\neq \{\ket{0},\ket{1}\}$,
$\bracket{\xi_0}{\xi_1}=0$, and $|\bracket{\xi_0}{0}|\neq
|\bracket{\xi_1}{0}|$.  We then have $S(X|B)=0$ and $S(Y|A)>0$ while
$X=\dep{X}{Y}$ and $Y=\dep{Y}{X}$.

To illustrate this definition of triviality, we argue in the following
that if a primitive $P_{X,Y}$ has a trivial regular embedding, there
exists a classical protocol which generates $X,Y$ securely in the HBC
model.  Let $\ket{\psi}\in \emb{P_{X,Y}}$ be trivial and assume
without loss of generality that $S(\dep{Y}{X}|A)=0$.  Intuitively,
this means that Alice can learn everything possible about Bob's
outcome $Y$ ($Y$ could include some private coin-flips on Bob's side,
but that is ``filtered out'' by the dependent part). More precisely,
Alice holding register $A$ can measure her part of the shared state to
completely learn a realization of $\dep{Y}{X}$, specifying
$P_{X|Y=y}$. 
She then chooses $X$ according to the distribution $P_{X|Y=y}$.  An
equivalent way of trivially generating $(X,Y)$ classically is the
following classical protocol:
\begin{enumerate}
\item Alice samples $P_{X|Y=y'}$ from distribution $P_{\dep{Y}{X}}$ and
  announces its outcome to Bob. She samples $x$ from the distribution
  $P_{X|Y=y'}$.
\item Bob picks $y$ with probability $P_{Y|\dep{Y}{X}=P_{X|Y=y'}}$ \, .
\end{enumerate}
Of course, the same reasoning applies in case $S(\dep{X}{Y}|B)=0$ with
the roles of Alice and Bob reversed. 

In fact, the following lemma (\append{proven in
  Appendix~\ref{app:triviality}}{whose proof can be found in the full
  version~\cite{SSS09arxiv}}) shows that any non-trivial primitive
$P_{X,Y}$ has a non-trivial embedding, i.e. there exists a quantum
protocol correctly implementing $P_{X,Y}$ while leaking less
information to QHBC adversaries than any classical protocol for
$P_{X,Y}$ in the HBC model.
\begin{lemma}
\label{nontrivialemb}
If $P_{X,Y}$ is a non-trivial primitive then the canonical embedding
$\ket{\psi_{\vec{0}}}\in \emb{P_{X,Y}}$ is also non-trivial.
\end{lemma}

\section{The Leakage of Quantum Embeddings}\label{crypto}

We formally define the leakage of embeddings and establish properties
of the leakage. 
The proofs of all statements in this section can be found in
\append{Appendix~\ref{app:crypto_with_emb}}{the full
  version~\cite{SSS09arxiv}}.

\subsection{Definition and Basic Properties of Leakage}\label{sleakage}
A perfect implementation of $P_{X,Y}$ simply provides $X$ to Alice and
$Y$ to Bob and does nothing else. The expected amount of information
that one random variable gives about the other is $I(X;Y)=
H(X)-H(X|Y)=H(Y)-H(Y|X) = I(Y;X)$. Intuitively, we define the {\em
  leakage of a quantum embedding $\ket{\psi}_{ABA'B'}$ of $P_{X,Y}$}
as the larger of the two following quantities: the extra amount of
information Bob's quantum registers $BB'$ provide about $X$ and the
extra amount Alice's quantum state in $AA'$ provides about $Y$
respectively in comparison to ``the minimum amount''
$I(X;Y)$.\footnote{\label{foot:guess}There are other natural
  candidates for the notion of leakage such as the difference in
  difficulty between guessing Alice's output $X$ by measuring Bob's
  final quantum state $B$ and based on the output of the ideal
  functionality $Y$. While such definitions do make sense, they turn
  out not to be as easy to work with and it is an open question
  whether the natural properties described later in this section can
  be established for these notions of leakage as well.}

\begin{definition}
Let $\ket{\psi}\in \hil_{ABA'B'}$ be an embedding of $P_{X,Y}$. 
We define the leakage $\ket{\psi}$ as
$$\Delta_{\psi}(P_{X,Y}):=\max \left\{ S(X;BB')-I(X;Y) \, , \, S(AA';Y)-I(X;Y) \right\}\, .$$
Furthermore, we say that $\ket{\psi}$ is 
{\em $\delta$-leaking} if $\Delta_{\psi}(P_{X,Y})\geq\delta$ .
\end{definition}

It is easy to see that the leakage is non-negative since $S(X;BB')\geq
S(X;\tilde{B})$ for $\tilde{B}$ the result of a quantum operation
applied to $BB'$.  Such an operation could be the trace over the extra
working register $B'$ and a measurement in the computational basis of
each qubit of the part encoding $Y$, yielding
$S(X;\tilde{B})=I(X;Y)$.

We want to argue that our notion of leakage is a good measure for the
privacy of the player's outputs. In the same spirit, we will argue
that the minimum achievable leakage for a primitive is related to the
``hardness'' of implementing it. We start off by proving several basic
properties about leakage.

For a general state in $\hil_{ABA'B'}$ the quantities
$S(X;BB')-I(X;Y)$ and $S(AA';Y)-I(X;Y)$ are not necessarily
equal. Note though that they coincide for regular embeddings
$\ket{\psi}\in\emb{P_{X,Y}}$ produced by a correct protocol (where the
work spaces $A'$ and $B'$ are empty): Notice that $ S(X;B) =
S(X)+S(B)-S(X,B) = H(X)+S(B) - H(X) = S(B)$ and because $\ket{\psi}$
is pure, $S(A)=S(B)$. Therefore, $S(X;B)=S(A;Y)$ and the two
quantities coincide. The following lemma states that this actually
happens for \emph{all} embeddings and hence, the definition
of leakage is symmetric with respect to both players.

\begin{lemma}[Symmetry]
\label{symmetry}
Let $\ket{\psi}\in \hil_{ABA'B'}$ be an embedding of
$P_{X,Y}$. Then,
$$\Delta_\psi(P_{X,Y})=S(X;BB')-I(X;Y)=S(AA';Y)-I(X;Y)\, .$$
\end{lemma}

The next lemma shows that the leakage of an embedding of a given 
primitive is lower-bounded by the leakage of some regular embedding 
of the same primitive, which simplifies the calculation of lower bounds for the leakage of embeddings.

\begin{lemma}
\label{super_leak}
For every embedding $\ket{\psi}$ of a primitive $P_{X,Y}$, there is a
regular embedding $\ket{\psi'}$ of $P_{X,Y}$ such that
$\Delta_\psi(P_{X,Y})\geq \Delta_{\psi'}(P_{X,Y}).$
\end{lemma}

So far, we have defined the leakage of an embedding of a primitive. The natural definition of 
the leakage of a primitive is the following.
\begin{definition}
\label{infimum}
We define the \emph{leakage of a primitive $P_{X,Y}$} as the minimal
leakage among all protocols correctly implementing
$P_{X,Y}$. Formally, 
$$\Delta_{P_{X,Y}} \assign \min_{\ket{\psi}}\Delta_{\psi}(P_{X,Y}) \, ,$$
where the minimization is over all embeddings $\ket{\psi}$ of $P_{X,Y}$.
\end{definition}
Notice that the minimum in the previous definition is well-defined,
because by Lemma~\ref{super_leak}, it is sufficient to minimize over regular embeddings
$\ket{\psi}\in\emb{P_{X,Y}}$. Furthermore, the function
$\Delta_{\psi}(P_{X,Y})$ is continuous on the compact (i.e.~closed and
bounded) set $[0,2\pi]^{|\mathcal{X}\times\mathcal{Y}|}$ of complex
phases corresponding to elements $\ket{x,y}_{AB}$ in the formula for
$\ket{\psi}_{AB}\in \emb{P_{X,Y}}$ and therefore it achieves its
minimum.

The following theorem shows that the leakage of any embedding of a
primitive $P_{X,Y}$ is lower-bounded by the minimal leakage achievable
for primitive $P_{\dep{X}{Y},\dep{Y}{X}}$ (which due to
Lemma~\ref{super_leak} is achieved by a regular embedding).
\begin{theorem}
\label{specialform2}
For any primitive $P_{X,Y}$, 
$\Delta_{P_{X,Y}} \geq \Delta_{P_{\dep{X}{Y},\dep{Y}{X}}}.$
\end{theorem}

\begin{proof}[Sketch]
  The proof idea is to pre-process the registers storing $X$ and $Y$
  in a way allowing Alice and Bob to convert a regular embedding of
  $P_{X,Y}$ (for which the minimum leakage is achieved) into a regular
  embedding of $P_{\dep{X}{Y},\dep{Y}{X}}$ by measuring parts of these
  registers.  It follows that on average, the leakage of the resulting
  regular embedding of $P_{\dep{X}{Y},\dep{Y}{X}}$ is at most the
  leakage of the embedding of $P_{X,Y}$ the players started
  with. Hence, there must be a regular embedding of
  $P_{\dep{X}{Y},\dep{Y}{X}}$ leaking at most as much as the best
  embedding of $P_{X,Y}$. See
  \append{Appendix~\ref{app:specialform}}{\cite{SSS09arxiv}} for the complete proof. \qed
\end{proof}

\subsection{Leakage as Measure of Privacy and Hardness of
  Implementation} \label{sec:limitedresources} The main results of
this section are consequences of the Holevo bound
(Theorem~\ref{holevo}).
\begin{theorem} \label{thm:nonleaktrivial} 
If a two-party quantum protocol provides the correct outcomes of
$P_{X,Y}$ to the players without leaking extra information, then $P_{X,Y}$
must be a trivial primitive.
\end{theorem}
\begin{proof} Theorem~\ref{specialform2} implies that if there is a
  $0$--leaking embedding of $P_{X,Y}$ than there is also a
  $0$--leaking embedding of $P_{\dep{X}{Y},\dep{Y}{X}}$. Let us
  therefore assume that $\ket{\psi}$ is a non-leaking embedding of
  $P_{X,Y}$ such that $X=\dep{X}{Y}$ and $Y=\dep{Y}{X}$.  We can write
  $\ket{\psi}$ in the form
  $\ket{\psi}=\sum_{x}\sqrt{P_X(x)}\ket{x}\ket{\varphi_x}$ and get
  $\rho_B = \sum_x P_X(x)\proj{\varphi_x}$. For the leakage of
  $\ket{\psi}$ we have:
  $\Delta_\psi(P_{X,Y})=S(X;B)-I(X;Y)=S(\rho_B)-I(X;Y)=0$. From the
  Holevo bound (Theorem~\ref{holevo}) follows that the states
  $\{\ket{\varphi_x}\}_x$ form an orthonormal basis of their span
  (since $X=\dep{X}{Y}$, they are all different) and that $Y$ captures
  the result of a measurement in this basis, which therefore is the
  computational basis. Since $Y=\dep{Y}{X}$, we get that for each $x$,
  there is a single $y_x\in\mathcal{Y}$ such that
  $\ket{\varphi_x}=\ket{y_x}$.
The primitives $P_{\dep{X}{Y},\dep{Y}{X}}$ and $P_{X,Y}$ are therefore
trivial. \qed
\end{proof}

In other words, the only primitives that two-party quantum protocols
can implement correctly (without the help of a trusted third party)
and without leakage are the trivial ones! We note that it is not
necessary to use the strict notion of correctness from
Definition~\ref{defcorrect} in this theorem, but a more complicated
proof can be done solely based on the correct distribution of the
values. This result can be seen as a quantum extension of the
corresponding characterization for the cryptographic power of
classical protocols in the HBC model. Whereas classical two-party
protocols cannot achieve anything non-trivial, their quantum
counterparts necessarily leak information when they implement
non-trivial primitives.

The notion of leakage can be extended to protocols involving a trusted
third party (see
\append{Appendix~\ref{app:tripartite}}{\cite{SSS09arxiv}}). A special
case of such protocols are the ones where the players are allowed one
call to a black box for a certain non-trivial primitive. It is natural
to ask which primitives can be implemented without leakage in this
case.  As it turns out, the monotones $H(\dep{X}{Y}|Y)$ and
$H(\dep{Y}{X}|X)$, introduced in~\cite{WW04}, are also monotones for
quantum computation, in the sense that all joint random variables
$X',Y'$ that can be generated by quantum players without leakage using
one black-box call to $P_{X,Y}$ satisfy $H(\dep{X'}{Y'}|Y')\leq
H(\dep{X}{Y}|Y)$ and $H(\dep{Y'}{X'}|X')\leq
H(\dep{Y}{X}|X)$. 

\begin{theorem}
\label{no-amplif}
Suppose that primitives $P_{X,Y}$ and $P_{X',Y'}$ satisfy $H(\dep{X'}{Y'}|Y')>H(\dep{X}{Y}|Y)$ 
or $H(\dep{Y'}{X'}|X')>H(\dep{Y}{X}|X)$. Then any implementation of $P_{X',Y'}$ using just one call 
to the ideal functionality for $P_{X,Y}$ leaks information. 
\end{theorem}

\subsection{Reducibility of Primitives and Their Leakage}
This section is concerned with the following question: Given two
primitives $P_{X,Y}$ and $P_{X',Y'}$ such that $P_{X,Y}$ is reducible
to $P_{X',Y'}$, what is the relationship between the leakage of
$P_{X,Y}$ and the leakage of $P_{X',Y'}$?  We use the notion of
reducibility in the following sense: We say that a primitive $P_{X,Y}$
is \emph{reducible in the HBC model} to a primitive $P_{X',Y'}$ if
$P_{X,Y}$ can be securely implemented in the HBC model from (one call
to) a secure implementation of $P_{X',Y'}$. The above question can
also be generalized to the case where $P_{X,Y}$ can be computed from
$P_{X',Y'}$ only with certain probability.  Notice that the answer,
even if we assume perfect reducibility, is not captured in our
previous result from Lemma~\ref{super_leak}, since an embedding of
$P_{X',Y'}$ is not necessarily an embedding of $P_{X,Y}$ (it might
violate the correctness condition). However, under certain
circumstances, we can show that $\Delta_{P_{X',Y'}}\geq
\Delta_{P_{X,Y}}.$
\begin{theorem}
\label{reduction}
Assume that primitives $P_{X,Y}$ and $P_{X',Y'}=P_{X'_0X'_1,Y'_0Y'_1}$ satisfy the condition:
$$
\sum_{x,y:P_{X'_0,Y'_0|X'_1=x,Y'_1=y}\simeq P_{X,Y}} P_{X'_1,Y'_1}(x,y)\geq 1-\delta,
$$
where the relation $\simeq$ means that the two distributions are equal
up to relabeling of the alphabet. Then, $\Delta_{P_{X',Y'}}\geq
(1-\delta)\Delta_{P_{X,Y}}.$
\end{theorem}
This theorem allows us to derive a lower bound on the leakage of
1-out-of-2 Oblivious Transfer of $r$-bit strings in
Section~\ref{sec:primleakage}.

\section{The Leakage of Universal Cryptographic
  Primitives}\label{sec:primleakage} In this section, we exhibit lower
bounds on the leakage of some universal two-party primitives, see
Appendix~\ref{app:primitives} for an overview of these primitives. In
the following table, $\srot$ denotes the $r$-bit string version of
randomized Rabin OT, where Alice receives a random $r$-bit string and
Bob receives the same string or an erasure symbol, each with
probability 1/2.  Similarly, $\sotr$ denotes the string version of
\ot, where Alice receives two $r$-bit strings and Bob receives one of
them. By $\otn$ we denote the noisy version of \ot, where the \ot\
functionality is implemented correctly only with probability $1-p$.
Table~\ref{tab:lowerbounds} summarizes the lower bounds on the leakage
of these primitives (the derivations can be found in
Appendix~\ref{app:universalleakage}). We note that Wolf and
Wullschleger~\cite{WW05b} have shown that a randomized \ot\ can be
transformed by local operations into an additive sharing of an AND
(here called \sand).  Therefore, our results for \ot\ below also apply
to \sand.

\begin{table}[h] \center \renewcommand{\arraystretch}{1.5}
\begin{tabular}{@{}l|c|@{\ \ }l@{}} \hline
{\bf primitive} & {\bf leaking at least} & {\bf comments} \\ \hline
$\srotp{1}$ & $(h(\frac{1}{4})-\frac{1}{2}) \approx 0.311$ & same
leakage for all regular embeddings \\ \hline 
$\srot$ & $(1-O(r2^{-r}))$ & same leakage for all regular embeddings\\ \hline
$\ot,\sand$ \hspace{2mm} & $\frac12$ & minimized by canonical embedding\\ \hline 
$\sotr$ & $(1-O(r2^{-r}))$ & (suboptimal) lower bound \\ \hline 
$\otn$ & $\frac{\left(1/2-p-\sqrt{p(1-p)}\right)^2}{8\ln 2}$ \hspace{2mm} &
if $p<\sin^2(\pi/8)\approx 0.15$, (suboptimal) lower bound \\ \hline
\end{tabular}
\medskip
\caption{Lower bounds on the leakage for universal two-party
  primitives} \label{tab:lowerbounds}
\vspace{-0.6cm}
\end{table}

\sotr\ and \otn\ are primitives where the direct evaluation of the
leakage for a general embedding $\ket{\psi_\theta}$ is hard, because
the number of possible phases increases exponentially in the number of
qubits. Instead of computing $S(A)$ directly, we derive (suboptimal)
lower bounds on the leakage.

Based on the examples of \srot\ and \ot, it is tempting to conjecture
that the leakage is always minimized for the canonical embedding,
which agrees with the geometric intuition that the minimal pairwise
distinguishability of quantum states in a mixture minimizes the von
Neumann entropy of the mixture. However, Jozsa and Schlienz have shown
that this intuition is sometimes incorrect~\cite{JS00}. In a quantum
system of dimension at least three, we can have the following
situation: For two sets of pure states $\{\ket{u_i}\}_{i=1}^n$ and
$\{\ket{v_i}\}_{i=1}^n$ satisfying $|\bracket{u_i}{u_j}|\leq
|\bracket{v_i}{v_j}|$ for all $i,j$, there exist probabilities $p_i$
such that for $\rho_u \assign \sum_{i=1}^np_i\ketbra{u_i}{u_i}$,
$\rho_v \assign \sum_{i=1}^np_i\ketbra{v_i}{v_i}$, it holds that
$S(\rho_u)<S(\rho_v)$.  As we can see, although each pair $\ket{u_i}$,
$\ket{u_j}$ is more distinguishable than the corresponding pair
$\ket{v_i}$, $\ket{v_j}$, the overall $\rho_u$ provides us with less
uncertainty than $\rho_v$. It follows that although for the canonical
embedding $\ket{\psi_{\vec{0}}}=\sum_y\ket{\varphi_y}\ket{y}$ of
$P_{X,Y}$ the mutual overlaps $|\bracket{\varphi_y}{\varphi_{y'}}|$
are clearly maximized, it does not necessarily imply that $S(A)$ in
this case is minimal over $\emb{P_{X,Y}}$. It is an interesting open
question to find a primitive whose canonical embedding does not
minimize the leakage or to prove that no such primitive exists.

For the primitive \potn, our lower bound on the leakage only holds for
$p<\sin^2(\pi/8)\approx 0.15$. Notice that in reality, the leakage is
strictly positive for any embedding of \potn\ with $p<1/4$, since for
$p<1/4$, \potn\ is a non-trivial primitive. On the other hand, $\potnq$ is a trivial primitive implemented securely by the following
protocol in the classical HBC model:
\begin{enumerate}
\item Alice chooses randomly between her input bits $x_0$ and $x_1$ and sends the chosen value $x_a$ to Bob. 
\item Bob chooses his selection bit $c$ uniformly at random and sets $y\assign x_a$.
\end{enumerate}
Equality $x_c=y$ is satisfied if either $a=c$, which happens with
probability $1/2$, or if $a\neq c$ and $x_a=x_{1-a}$, which happens
with probability $1/4$. Since the two events are disjoint, it follows
that $x_c=y$ with probability $3/4$ and that the protocol implements
\potnq. The implementation is clearly secure against
honest-but-curious Alice, since she does not receive any message from
Bob. It is also secure against Bob, since he receives only one bit
from Alice. By letting Alice randomize the value of the bit she is
sending, the players can implement \potn\ securely for any value
$1/4<p\leq 1/2$.

\section{Conclusion and Open Problems}\label{conclusion}
We have provided a quantitative extension of qualitative impossibility
results for two-party quantum cryptography. All non-trivial primitives
leak information when implemented by quantum protocols.  Notice that
demanding a protocol to be non-leaking does in general not imply the
privacy of the players' outputs.  For instance, consider a protocol
implementing \ot\ but allowing a curious receiver with probability
$\frac12$ to learn both bits simultaneously or with probability
$\frac12$ to learn nothing about them.  Such a protocol for \ot\ would
be non-leaking but nevertheless insecure.  Consequently,
Theorem~\ref{thm:nonleaktrivial} not only tells us that any quantum
protocol implementing a non-trivial primitive must be insecure, but
also that a privacy breach will reveal itself as
leakage. Our framework allows to quantify the leakage of any two-party quantum
protocol correctly implementing a primitive. The impossibility results
obtained here are stronger than standard ones since they only rely on
the cryptographic correctness of the protocol. Furthermore, we present
lower bounds on the leakage of some universal two-party primitives.

A natural open question is to find a way to identify good embeddings
for a given primitive. In particular, how far can the leakage of the
canonical embedding be from the best one? Such a characterization,
even if only applicable to special primitives, would allow to lower
bound their leakage and would also help to understand the power of
two-party quantum cryptography in a more concise way.

It would also be interesting to find a measure of cryptographic
non-triviality for two-party primitives and to see how it relates to
the minimum leakage of any implementation by quantum protocols. For
instance, is it true that quantum protocols for primitive $P_{X,Y}$
leak more if the minimum (total variation) distance between $P_{X,Y}$
and any trivial primitive increases?  

Another question we leave for future research is to define and
investigate other notions of leakage, e.g.~in the one-shot setting
instead of in the asymptotic regime (as outlined in
Footnote~\ref{foot:guess}). Results in the one-shot setting have
already been established for data compression~\cite{RW05}, channel
capacities~\cite{RWW06}, state-merging~\cite{WR07,Berta08} and other
(quantum-) information-theoretic tasks.

Furthermore, it would be interesting to find more applications for the
concept of leakage, considered also for protocols using an environment
as a trusted third party. In this direction, we have shown in
Theorem~\ref{no-amplif} that any two-party quantum protocol for a
given primitive, using a black box for an ``easier'' primitive, leaks
information. Lower-bounding this leakage is an interesting open
question. We might also ask how many copies of the ``easier''
primitive are needed to implement the ``harder'' primitive by a
quantum protocol, which would give us an alternative measure of
non-triviality of two-party primitives.

\bibliographystyle{alpha}

\bibliography{crypto,qip,procs}

\append{
\appendix

\section{Cryptographic Primitives}
\label{app:primitives}
Here we list the standard cryptographic primitives studied in this paper.

\begin{description}
\item[String Rabin OT ($\srot$):]~\cite{Rabin81} Alice sends a random
  string of $r$ bits to Bob who receives it with probability $1/2$,
  otherwise he receives a special symbol $\bot$. Alice does not learn
  any information about whether Bob has received the string she sent.

\item[One-out-of-two String OT ($\sotr$):]~\cite{Wiesner83,EGL82}
Alice sends two random $r$-bit strings to Bob who decides which of them he receives. Bob does not learn 
any information about the other one of Alice's strings and Alice does not learn which of the strings has been received by Bob.

\item[Additive sharing of AND ($\sand$):]~\cite{PR94}
Alice and Bob choose their respective input bits $x$ and $y$, and receive the output bits $a$ resp. $b$ such that $a\oplus b=x\wedge y$ and $\Pr[a=0]=1/2$. They do not get any other information.

\item[Noisy one-out-of-two OT ($\otn$):] Alice sends two bits to Bob
 who decides which of them he wants to receive. The selected bit is
 transmitted to him over a noisy channel with noise rate $p$. Bob
 does not learn any information about the other one of Alice's bits
 and Alice does not learn any information about Bob's selection bit.
\end{description}

We present a description of the randomized versions of the primitives
in the following: 

\begin{description}
\item[String Rabin OT ($\srot$):]
For $x\in\{0,1\}^r$ and $y\in\{0,1\}^r\cup \{\bot\}$:
\[\psrot(x,y)=\left\{\begin{array}{ll}{2^{-r-1}} &\mbox{if $x=y$ or $y=\bot$,}\\
                                    0                 &\mbox{otherwise},
\end{array}\right.
\]
is the joint probability distribution associated to an execution of
Rabin OT of a random binary string of length $r$.

\item[One-out-of-two OT (\ot):]
For $x_0,x_1,y,c\in\{0,1\}$:
\[\pot((x_0,x_1),(c,y))=
\left\{\begin{array}{ll}\frac{1}{8} & \mbox{if $y=x_c$,}\\
                             0       & \mbox{otherwise}, 
                          \end{array}\right.
\]
is the joint probability distribution for the execution of one-out-of-two OT
upon random input bits.
\item[One-out-of-two String OT ($\sotr$):]
For $x_0,x_1,y\in\{0,1\}^r$ and $c\in\{0,1\}$, let
\[\psot((x_0,x_1),(c,y))=\left\{\begin{array}{ll}{2^{-2r-1}} &\mbox{if $y=x_c$,}\\
                                                            0    &\mbox{otherwise},
\end{array}\right.
\]
is the joint probability distribution associated to an execution of
one-out-of-two $r$-bit string OT upon random inputs.
\item[Additive Sharing of AND (\sand):] For $x,y,a,b\in\{0,1\}$:
\[\pnl((x,a),(y,b))=\left\{\begin{array}{ll}\frac{1}{8} & \mbox{if $xy=a\oplus b$,}\\
                                                 0       & \mbox{otherwise}, 
                          \end{array}\right.
\]
is the joint probability distribution associated to the generation of an additive
sharing for the {\sc and} of two random bits.

\item[Noisy one-out-of-two OT ($\otn$):]
For $x_0,x_1,y,c\in\{0,1\}$ and $p\in\,(0,1/2)$:
\[\potn((x_0,x_1),(c,y))=\left\{\begin{array}{ll}\frac{1-p}{8} &\mbox{if $y=x_c$,}\\
                                                \frac{p}{8} &\mbox{otherwise},
\end{array}\right.
\]
is the joint probability distribution associated to an execution
of one-out-of-two OT where the selected bit is received  through a binary symmetric
channel with error rate $p$.
\end{description}

\section{Proof of Lemma~\ref{nontrivialemb}} \label{app:triviality}
A non-trivial embedding of $P_{X,Y}$ can be created from a non-trivial embedding of  $P_{\dep{X}{Y},\dep{Y}{X}}$ by applying local unitary transforms.  We therefore assume
without loss of generality that $X=\dep{X}{Y}$ and $Y=\dep{Y}{X}$.
Let
$$\ket{\psi_{\vec{0}}} \assign \sum_{x,y}\sqrt{P_{X,Y}(x,y)}\ket{x,y}$$ be the canonical
embedding of $P_{X,Y}$.
Since $X=\dep{X}{Y}$ and $Y=\dep{Y}{X}$, it holds for any
$x_0\neq x_1$ that $P_{Y|X=x_0}\neq P_{Y|X=x_1}$. Furthermore, since $P_{X,Y}$
is non-trivial, there exist $x_0\neq x_1$ and $y_0$ such that $P_{Y|X=x_0}(y_0)>0$
and $P_{Y|X=x_1}(y_0)>0$.
The state $\ket{\psi_{\vec{0}}}$ can be written in the form:
\begin{eqnarray*}
\ket{\psi_{\vec{0}}}&=&\sqrt{P_{X}(x_0)}\ket{x_0}\sum_{y}\sqrt{P_{Y|X=x_0}(y)}\ket{y}+\sqrt{P_{X}(x_1)}\ket{x_1}\sum_{y}\sqrt{P_{Y|X=x_1}(y)}\ket{y}+\ket{\psi'},
\end{eqnarray*}
where
$\tr(\proj{x_0}\tr_B\proj{\psi'})=\tr(\proj{x_1}\tr_B\proj{\psi'})=0$.
Set $\ket{\varphi^{x_b}}\assign \sum_y \sqrt{P_{Y|X=x_b}(y)}\ket{y}$ for $b\in\{0,1\}$.
Since $P_{Y|X=x_0}\neq P_{Y|X=x_1}$, we get that
$|\bracket{\varphi^{x_0}}{\varphi^{x_1}}|<1$.  Because all coefficients
at $\ket{y}$ in the normalized vectors $\ket{\varphi^{x_0}}$ and
$\ket{\varphi^{x_1}}$  are non-negative, and the coefficients at
$\ket{y_0}$ are both positive,
$\bracket{\varphi^{x_0}}{\varphi^{x_1}}\neq 0$. Therefore, the
non-identical states $\ket{\varphi^{x_0}}$ and $\ket{\varphi^{x_1}}$
cannot be perfectly distinguished, which implies that Bob cannot learn
whether $X=x_0$ or $X=x_1$ with probability 1. Therefore, the von
Neumann entropy on Bob's side $S(B)$ is such that $S(B)<H(X)$. As
$H(\dep{X}{Y}|Y)>0$ implies $H(\dep{Y}{X}|X)>0$, we can argue in the
same way that $S(A)<H(Y)$ from which follows that
$\ket{\psi_{\vec{0}}}$ is a non-trivial quantum embedding of
$P_{X,Y}$.  \qed

\section{Proofs of Properties of Leakage}
\label{app:crypto_with_emb}
 
\subsection{Proof of Lemma~\ref{symmetry}}
We have already shown that the statement is true in the case where both $A'$ 
and $B'$ are trivial. In the case where $A'$ is trivial and
$B'$ is not, the Markov chain condition 
implies that $\ket{\psi}$ is of the form
$$\ket{\psi}=\sum_{x,y}\sqrt{P_{X,Y}(x,y)}\ket{x,y}_{AB}\ket{\varphi^y}_{B'}\, ,$$
hence, Bob can fix $y_0$ and apply a unitary transform $U_{BB'}$
on his part of the system,
such that $U_{BB'}\ket{y,\varphi^y}=\ket{y,\varphi^{y_0}}$,
and
$$\id_A\otimes U_{BB'}\ket{\psi}_{ABB'}=\ket{\psi^*}_{AB}\otimes \ket{\varphi^{y_0}}_{B'}\, ,$$
where $\ket{\psi^*}\in\emb{P_{X,Y}}$. In the resulting product state, 
$S(X;BB')-I(X;Y)=S(X;B)-I(X;Y)=S(A;Y)-I(X;Y)$, due to the fact that 
$\ket{\psi^*}\in\emb{P_{X,Y}}$. An analogous statement holds
in the case where $B'$ is trivial and $A'$ is non-trivial.

We now assume that both $A'$ and $B'$ are non-trivial. 
An embedding of $P_{X,Y}$ can be written as $ \ket{\psi} =
\sum_{x,y} \sqrt{P_{X,Y}(x,y)} \ket{x,y}_{AB}
\ket{\varphi^{x,y}}_{A'B'}$.

For every $x$ and $y$, we can write the pure
state 
$$\ket{\varphi^{x,y}}_{A'B'} =
\sum_{k=1}^K \sqrt{\lambda_k^{x,y}} \ket{e_k^{x,y}}_{A'} \ket{f_k^{x,y}}_{B'}$$ 
in Schmidt form. For the reduced density matrices, we obtain
$$\rho_{A'}^{x,y} = \sum_k \lambda_k^{x,y} \proj{e_k^{x,y}}\, .$$

Since any embedding $\ket{\psi}\in \hil_{ABA'B'}$ of $P_{X,Y}$ is
produced by a correct protocol, it satisfies
$$S(XA';B)=S(X;YB')=I(X;Y)$$ 
which is
equivalent to $A'\leftrightarrow X\leftrightarrow Y$ and
$X\leftrightarrow Y\leftrightarrow B'$ being Markov chains. It follows
that for every $x$ and $y \neq y'$, the reduced density matrices
$\rho_{A'}^{x,y}=\rho_{A'}^{x,y'}=\rho_{A'}^x$ coincide and therefore,
the eigenvalues $\lambda_k^{x,y}$ cannot depend on $y$. Because of
$X\leftrightarrow Y\leftrightarrow B'$, they can neither depend on
$x$. Hence, $\ket{\varphi^{x,y}} = \sum_k \sqrt{\lambda_k}e^{i\theta'(k,x,y)}
\ket{e_k^x}\ket{f_k^y}$. The phase factors arise from the fact
that from a reduced density matrix the global phases of the
Schmidt-basis elements cannot be determined.

Let us fix a set of orthogonal states $\{\ket{k}\}_k$. We define the unitary $U_{AA'}$ to be the mapping of
the orthonormal states $\{\ket{e_k^{x}}\}_k$ into the orthonormal
states $\{\ket{k}\}_k$. Note that $U_{AA'}$ only acts on
register $A'$ conditioned on the $x$-value in $A$. Analogously, let
$U_{BB'}$ map the states $\{\ket{f_k^{y}}\}_k$ into
$\{\ket{k}\}_k$. Applying $U_{AA'} \otimes U_{BB'}$ to
$\ket{\psi}$ results into state
\begin{eqnarray*}
&\sum_{x,y} \sqrt{P_{X,Y}(x,y)} \ket{x,y}_{AB} \sum_k
\sqrt{\lambda_k} e^{i\theta'(k,x,y)}\ket{k,k}_{A'B'}\\
= &\sum_k \sqrt{\lambda_k} \left(\sum_{x,y}\sqrt{P_{X,Y}(x,y)}
e^{i\theta'(k,x,y)}\ket{x,y}\right) \ket{k,k}\\
= &\sum_k \sqrt{\lambda_k} \ket{\psi_k}_{AB} \otimes \ket{k,k}_{A'B'},
\end{eqnarray*}  
where each $\ket{\psi_k}_{AB}\in\emb{P_{X,Y}}$. 
The cqq-state $\rho_{XBB'}$ can now be written in the form:
$$\rho_{XBB'}=\sum_x P_X(x)\proj{x}\otimes\sum_k \lambda_k
\proj{\phi^x_k,k} \, ,$$
where $\ket{\phi^x_k}=\sum_y \sqrt{P_{Y|X=x}}e^{i\theta'(k,x,y)}\ket{y}$. 
Due to the second component, the states $\ket{\phi^x_k,k}$ are mutually 
orthogonal for each $x$. Therefore, for each $x$,  
$$S\left(\sum_k\lambda_k \proj{\phi^x_k,k}\right)=H(\lambda_1,\dots,\lambda_K)\, .$$
As a result we get that
$$S(XBB')=H(X)+\sum_x P_X(x)H(\lambda_1,\dots,\lambda_K)=H(X)+H(\lambda_1,\dots,\lambda_K)$$ 
and analogously, 
$$S(AA'Y)=H(Y)+H(\lambda_1,\dots,\lambda_K)\, ,$$
yielding the desired statement as follows:
\begin{eqnarray*}
S(X;BB')-I(X;Y)&= &H(X)+S(BB')-S(XBB')-I(X;Y)\\
&=& H(X)+S(BB')-(H(X)+H(\lambda_1,\dots,\lambda_K))-I(X;Y)\\
&=& H(Y)+S(AA')-(H(Y)+H(\lambda_1,\dots,\lambda_K))-I(X;Y)\\
&=& S(AA';Y)-I(X;Y).
\end{eqnarray*}
The equality $S(AA')=S(BB')$ follows from the purity of $\ket{\psi}$.
\qed

\subsection{Proof of Lemma~\ref{super_leak}}
In the case where $A'$ and $B'$ are both trivial, then
$\ket{\psi}\in\emb{P_{X,Y}}$ is a regular embedding and the statement
holds trivially. In the case where $A'$ is trivial and $B'$ is not, we
have shown in the proof of Lemma~\ref{symmetry} that an embedding
$\ket{\psi}$ of $P_{X,Y}$ is locally equivalent to a state
$\ket{\psi^*}\otimes\ket{\sigma}$ for $\ket{\psi^*}\in\emb{P_{X,Y}}$
and a pure state $\ket{\sigma}$.  An analogous statement holds if $B'$
is trivial and $A'$ is not.  Therefore, in these two cases we get for
some $\ket{\psi^*}\in\emb{P_{X,Y}}$ that
$\Delta_\psi=\Delta_{\psi^*}$.

Now assume that both $A'$ and $B'$ are non-trivial. 
Embedding $\ket{\psi}$ of $P_{X,Y}$ can be written as 

$ \ket{\psi} =
\sum_{x,y} \sqrt{P_{X,Y}(x,y)} \ket{x,y}_{AB}
\ket{\varphi^{x,y}}_{A'B'}$.

In the proof of Lemma~\ref{symmetry} we 
show the existence of two local unitary
transforms $U_{AA'}$ and $U_{BB'}$ on Alice's and Bob's side 
that
transform $\ket{\psi}$ into 
$\sum_k \sqrt{\lambda_k} \ket{\psi_k}_{AB} \otimes \ket{k,k}_{A'B'}$ 
for a set of orthogonal states $\{\ket{k}\}_k$ and $\ket{\psi_k}\in\emb{P_{X,Y}}$ 
for each $k$. 

If Alice measures register $A'$ or Bob measures $B'$ in the basis
$\{\ket{k}\}_k$, she/he transforms the state defined above into the
state $\ket{\psi_k}_{AB}\otimes\ket{k,k}_{A'B'}$ with probability
$\lambda_k$. Measuring register $A'$ arbitrarily does on average not
increase $S(AA';Y)$, and analogously, measuring $B'$ does not increase
$S(X;BB')$ on average. Hence, it follows from Holevo bound
(Theorem~\ref{holevo}) that 
$$ S(AA';Y)=S(A;Y)+S(A';Y|A)\geq
S(A;Y)+S(K;Y|A)=S(AK;Y) \, ,$$ 
where $K$ denotes the random variable
associated with the measurement of register $A'$ in the computational
basis.  Therefore, the leakage of $\ket{\psi}$ is at least the average
leakage of one particular strategy, i.e.~ $\Delta_{\psi}\geq
\sum_k \lambda_k \Delta_{\psi_k}$. Hence, there must exist a $k$ such that
for $\ket{\psi^*}:=\ket{\psi_k}$, it holds that $\Delta_{\psi}\geq
\Delta_{\psi^*}.$ \qed
 
\subsection{Proof of Theorem~\ref{specialform2}} \label{app:specialform}
In fact, the random variables $\dep{X}{Y}$ and $\dep{Y}{X}$ in the
claim can be replaced by any variables $X'$ and $Y'$, satisfying that
$X\leftrightarrow X'\leftrightarrow Y$ and $X\leftrightarrow
Y'\leftrightarrow Y$ are Markov chains, and that $Y'=f_Y(Y)$ and
$X'=f_X(X)$ for some deterministic functions $f_Y$ and $f_X$. For such
random variables we then have $I(X';Y')=I(X;Y)$. Therefore, showing
that for $\ket{\psi}\in\emb{P_{X,Y}}$ with the lowest leakage among all embeddings 
of $P_{X,Y}$ (its regularity follows from Lemma~\ref{super_leak}) 
and for some $\ket{\psi^*}\in\emb{P_{X',Y'}}$ , it holds that
$$S_\psi(B)-I(X;Y)=\Delta_\psi(P_{X,Y})\geq
\Delta_{\psi^*}(P_{X',Y'})=S_{\psi^*}(B)-I(X';Y')$$ 
is equivalent to proving $S_\psi(B)\geq S_{\psi*}(B)$. First, we show
that there exists $\ket{\tilde{\psi}}\in\emb{P_{X,Y'}}$ such that
$S_{\psi}(B)\geq S_{\tilde{\psi}}(B)$, i.e. $\Delta_\psi(P_{X,Y})\geq
\Delta_{\tilde{\psi}}(P_{X,Y'})$. The existence of $\ket{\psi^*}$ such
that $\Delta_{\tilde{\psi}}(P_{X,Y'})\geq \Delta_{\psi^*}(P_{X',Y'})$
follows from an analogous argument.

State $\ket{\psi}$ can be written in the form:
$$\ket{\psi}=\sum_{x,y}\sqrt{P_{X,Y}(x,y)}e^{i\theta(x,y)}\ket{x,y} \,
.$$

For any realization $y'$ of $Y'$, let $O_{y'} \assign \{y:\ f_{Y}(y)=y'\}$. WLOG assume that 
$O_{y'}=\{1,\dots,k_{y'}\}$. Let $g$ be a bijection of the form $g(y)=(f_{Y}(y),j_{y})$, where 
$j_{y}\in\{1,\dots,k_{f_{Y}(y)}\}$. 
A pair $(y',j)$ determines its $g$-preimage uniquely and therefore,  in the following we 
sometimes encode $y$ by $f_{Y}(y)j_{y}=(y',j)$. 
Formally, there is a unitary transform $U$ of Bob such that
\begin{eqnarray}
\id_A\otimes U\ket{\psi}\ket{0}_B&=&\sum_{x,y}\sqrt{P_{X,Y}(x,y)}e^{i\theta(x,y)}\ket{x,f_{Y}(y)j_{y}}\nonumber\\
&=&
\sum_{x,y'}\sqrt{P_{X,Y'}(x,y')}\ket{x,y'}\sum_{j=1}^{k_{y'}}\sqrt{P_{Y|Y'=y'}(g^{-1}(y',j))}e^{i\theta(x,g^{-1}(y',j))}\ket{j}
\, . \label{swap}
\end{eqnarray}

Our goal for the rest of the proof is to transform the register containing 
$j$ into a form 
where the order of the
summations over $(x,y')$ and $j$ in \eqref{swap} can be reversed 
to get a state of the
form
\[ \frac{1}{\sqrt{t}}\sum_{j=1}^t \ket{\hat{\psi}_j}_{AB}\ket{j}_B, \]
where $t$ is some normalization factor and each $\ket{\hat{\psi}_j}$ is in
$\emb{P_{X,Y'}}$.  Our claim that there
exists a state $\ket{\tilde{\psi}} \in \emb{P_{X,Y'}}$ such that
$S_{\tilde{\psi}}(B) \leq S_\psi(B)$ then follows from concavity of Von Neumann
entropy i.e., from the fact that the average of the entropies of the states
$\{\tr_A\proj{\hat{\psi}_j}\}_j$ is smaller than the entropy of their mixture 
which is equivalent to $\tr_A\proj{\psi}$.

In order to reverse the order of summation in \eqref{swap}, we show
that there exists a unitary $W$ on Bob's system such that
$$(\id_A\otimes W)(\id_A\otimes
U\ket{\psi}\ket{0}_B)\ket{0}_B = \ket{\varphi}=\frac{1}{\sqrt{t}}\sum_{z=1}^t \ket{\hat{\psi}_z}_{AB}\ket{z}_B\, ,$$
where each $\ket{\hat{\psi}_z}$ is a quantum embedding of a joint random variable $\hat{X}\hat{Y}$,
with the distribution arbitrarily close to distribution $P_{X,Y'}$.

Equality~(\ref{swap}) suggests to construct the states
$\ket{\hat{\psi}_z}_{AB}$ by disentangling the register containing $j$
from the registers containing $(x,y')$. This method will indeed lead us
to the result but only after some pre-processing of the register
containing $j$. First, we show how to split the register with $j$ for
each value of $y'$ into a uniform superposition of $t$ values which
Bob can measure afterwards to determine the index $z$ of an embedding
$\ket{\hat{\psi}_z}$. The uniformity over the register
containing the indices ensures that measuring the index does not have
any impact on the probability distribution $P_{X,Y'}$
implemented by $\ket{\hat{\psi}_z}$.

Consider $t\in\mathbb{N}$ such that
$0<1/t\ll\min_{y}\{P_{Y|Y'=f_{Y}(y)}(y)\}$. We can ensure that each $y'\in \mathcal{Y}'$ is split 
into 
exactly $t$ index-values $z$, by
adaptively defining a function $[tP_{Y|f_{Y}(Y)=y'}(y)]_{y}\in\{\lceil\ \rceil,\lfloor\ 
\rfloor\}$, indicating into how many values $z$ a given $y$ such that $f_{Y}(y)=y'$ splits. This procedure is elementary, but somewhat technical, and we postpone the detailed
description to the end of the proof.

For an event $y'$ of $Y'$, define $t_0 \assign 0$ and for $i\in\{1,\dots,k_{y'}\}$, 
$$t_i \assign \sum_{j\leq i}[tP_{Y|f_{Y}(Y)=y'}(y',j)]_{y',j}\, .$$ 
Let Bob's unitary transform $W$ acting upon the registers containing $Y'$, $j\in\{1,\dots,k_{Y'}\}$ and ancillas set to 0, be defined as follows:
$$W\ket{y',j}\ket{0}=\ket{y'}\frac{1}{\sqrt{[ P_{Y|f_{Y}(Y)=y'}(y',j)t]_{y',j}}}\sum_{z=t_{j-1}+1}^{t_{j}}\ket{z}\, .$$

The definition of $[\ ]_{y}$ implies that for each $y'$: $t_{k_{y'}}=t$, thus $z\in\{1,\dots,t\}$. We can write
\begin{eqnarray}
\ket{\varphi}& \assign &(\id_A\otimes W)((\id_A\otimes U)\ket{\psi}_{AB}\ket{0}_B)\ket{0}_B\nonumber\\
&=&\sum_{x,y'}\sqrt{P_{X,Y'}(x,y')}\ket{x,y'}\sum_{j=1}^{k_{y'}} \sqrt{\frac{P_{Y|f_{Y}(Y)=y'}(y',j)}{[ P_{Y|f_{Y}(Y)=y'}(y',j)t]_{y',j}}}e^{i\theta(x,(y',j))}\sum_{z=t_{j-1}+1}^{t_{j}}\ket{z}.\label{tz}
\end{eqnarray}

For the term $\frac{P_{Y|f_{Y}(Y)=y'}(y')}{[P_{Y|f_{Y}(Y)=y'}(y)t]_{y}}$ from~(\ref{tz}) we have
\begin{eqnarray}
\left|\frac{P_{Y|f_{Y}(Y)=y'}(y)}{[ P_{Y|f_{Y}(Y)=y'}(y)t]_{y}}-\frac{1}{t}\right|&=&
\left|\frac{tP_{Y|f_{Y}(Y)=y'}(y)-[ P_{Y|f_{Y}(Y)=y'}(y)t]_{y}}{t[ P_{Y|f_{Y}(Y)=y'}(y)t]_{y}}\right|\nonumber\\
&\leq& \frac{1}{P_{Y|f_{Y}(Y)=y'}(y)t^2-t}= \frac{1}{P_{Y|f_{Y}(Y)=y'}(y)t^2}+O\left(\frac{1}{t^3}\right).\label{fracaprox}
\end{eqnarray}
Now we can finally swap the summations to isolate $z$ as promised earlier. From~(\ref{tz}) and (\ref{fracaprox}) follows that
\begin{eqnarray*}
\ket{\varphi}&=&\sum_{x,y'}\sqrt{P_{X,Y'}(x,y')}\ket{x,y'}\sum_{z=1}^{t}e^{i\theta'(x,y',z)}\sqrt{\frac{1}{t}+\frac{\varepsilon(y',z)}{t^2}}\ket{z}\\
&=&\frac{1}{\sqrt{t}}\sum_{z=1}^{t}\left(\sum_{x,y'}e^{i\theta'(x,y',z)}\sqrt{1+\frac{\varepsilon(y',z)}{t}}\sqrt{P_{X,Y'}(x,y')}\ket{x,y'}\right)\ket{z},
\end{eqnarray*}
where $|\varepsilon(y',z)|\leq \frac{1}{\min_{y}{\{P_{Y|Y'=f_{Y}(y)}(y)}\}}$ and since a pair $(y',z)$ uniquely determines $y$ that it came from, 
$\theta'(x,y',z)=\theta(x,y)$ for $y$ corresponding to $(y',z)$.
If Bob measures $z$, the state $\ket{\varphi}$ collapses to
$$\left(\sum_{x,y'}e^{i\theta'(x,y',z)}\sqrt{1+\frac{\varepsilon(y',z)}{t}}\sqrt{P_{X,Y'}(x,y')}\ket{x,y'}\right)\otimes \ket{z}=\ket{\hat{\psi}_z}\otimes\ket{z}\, .$$
The state $\ket{\hat{\psi}_z}$ lies in $\emb{P_{\hat{X},\hat{Y}}}$ for a joint probability distribution  $P_{\hat{X},\hat{Y}}$ which is arbitrarily close to $P_{X,Y'}$. The distance of the two distributions depends on the choice of $t$.

Hence, for any  $\delta>0$ there is a way to pick a unitary transform
$W_\delta$ (with $t$ large enough) such that after applying $W_\delta$
and measuring $z$, the corresponding quantum systems satisfy
$|S_{\tilde{\psi}_z}(B)-S_{\hat{\psi}_z}(B)|\leq\delta$ for some
$\ket{\tilde{\psi}_z}\in\emb{P_{X,Y'}}$.

Concavity of Von Neumann entropy together with the fact that the state $\frac{1}{\sqrt{t}}\sum_{z=1}^t\ket{\hat{\psi}_z}\ket{z}$ is locally equivalent to $\ket{\psi}$ imply that
$$\frac{1}{t}\sum_{z=1}^tS_{\hat{\psi}_z}(B)\leq S_\psi(B)\, .$$
Therefore,  $S_\psi(B)\geq \min_z\{S_{\hat{\psi}_z}(B)\}$, and $S_\psi(B)\geq\min_z\{S_{\tilde{\psi}_z}(B)\}-\delta$ for $\delta$ arbitrarily small.

Continuity of Von Neumann entropy yields $S_\psi(B)\geq S_{\tilde{\psi}}(B)$ for some
$\ket{\tilde{\psi}}\in\emb{P_{X,Y'}}$, which is what we wanted to show.

Finally, it remains to give the correct definition of $[\ ]_{y}$:
For any $y'$, let us start by setting $[tP_{Y|f_{Y}(Y)=y'}(y)]_{y} \assign \lfloor tP_{Y|f_{Y}(Y)=y'}(y)\rfloor$ for
all $y: f_{Y}(y)=y'$. We now increase the value of
$[tP_{Y|f_{Y}(Y)=y'}(y)]_{y}$ in steps and show that at some point, this value equals $t$.
Let $0\leq i\leq k_{y'}$. In the $i$-th step, replace $[\ ]_{y',i}=\lfloor\ \rfloor$ with $[\ ]_{y',i}=\lceil\ \rceil$. After $k_{y'}$ steps, $[\ ]_{y}=\lceil\ \rceil$ for all $y: f_{Y}(y)=y'$. In every step the sum $\sum_{y, f_{Y}(y)=y'} [tP_{Y|f_{Y}(Y)=y'}(y)]_{y}$ increases by at most 1. Clearly, since $\sum_{y}P_{Y|f_{Y}(Y)=y'}(y)=1$, we get that
$$\sum_{y,f_{Y}(y)=y'}\lfloor tP_{Y|f_{Y}(Y)=y'}(y)\rfloor\leq t\ \ \  {\rm and}\ \ \ \sum_{y,f_{Y}(y)=y'}\lceil tP_{Y|f_{Y}(Y)=y'}(y)\rceil\geq t\, ,$$
thus for some $i$, $\sum_{y, f_{Y}(y)=y'} [tP_{Y|f_{Y}(Y)=y'}(y)]_{y}=t$.
\qed

\subsection{Proof of Theorem~\ref{reduction}}
State $\ket{\psi}_{A_0A_1B_0B_1}\in \emb{P_{X',Y'}}$ can be written in the form:
$$\ket{\psi}=\sum_{x\in\mathcal{X}'_1} \sqrt{P_{X'_1}(x)}\ket{x}_{A_1}\ket{\psi^x}_{A_0B}\, ,$$ 
where each
$\ket{\psi^x}$ 
is a regular embedding of $P_{X'_0Y'_0Y'_1|X'_1=x}$.  
Since 
$$S_\psi(Y'|A)\leq S_\psi(Y'|A_0,X'_1)=\sum_x P_{X'_1}(x)S_{\psi^x}(Y'|A_0,X'_1=x)\, ,$$
we obtain for the leakage of $\ket{\psi}$ that
\begin{eqnarray*}
\Delta_\psi(P_{X',Y'})&=&H(Y'|X')-S_\psi(Y'|A)\\
&\geq& H(Y'|X')-\sum_x P_{X'_1}(x)S_{\psi^x}(Y'|A_0,X'_1=x)\\
&=& \sum_x P_{X'_1}(x)(H(Y'|X'_0,X'_1=x)-S_{\psi^x}(Y'|A_0,X'_1=x))\\
&=& \sum_x P_{X'_1}(x)\Delta_{\psi^x}(P_{X'_0,Y'_0Y'_1|X'_1=x}) \, .
\end{eqnarray*}

By applying the same argument to each $\ket{\psi^x}$, we obtain that 
\begin{equation}
\label{reduc}
\Delta_\psi(P_{X',Y'})\geq
\sum_{xy}P_{X'_1,Y'_1}(x,y)\Delta_{\psi^{x,y}}(P_{X'_0,Y'_0|X'_1=x,Y'_1=y})
\, ,
\end{equation}
where each $\ket{\psi^{x,y}}$ is a regular embedding of $P_{X'_0,Y'_0|X'_1=x,Y'_1=y}$. 
For each $(x,y)$ such that 
$P_{X'_0,Y'_0|X'_1=x,Y'_1=y}\simeq P_{X,Y}$ is satisfied, we get that 
$$\Delta_{\psi^{x,y}}(P_{X'_0,Y'_0|X'_1=x,Y'_1=y})\geq \Delta_{P_{X,Y}}\, .$$ 
Since $\sum_{x,y:P_{X'_0,Y'_0|X'_1=x,Y'_1=y}\simeq P_{X,Y}} P_{X'_1,Y'_1}(x,y)\geq 1-\delta,$ 
we get from~(\ref{reduc}) that 
$$\Delta_\psi(P_{X',Y'})\geq (1-\delta)P_{X,Y}\, .$$ 
\qed

\subsection{Proof of Theorem~\ref{thm:nonleaktrivial}}
Theorem~\ref{specialform2} implies that if there is a $0$--leaking
embedding of $P_{X,Y}$ than there is also a $0$--leaking embedding of
$P_{\dep{X}{Y},\dep{Y}{X}}$. Let us therefore assume that $\ket{\psi}$
is a non-leaking embedding of $P_{X,Y}$ such that $X=\dep{X}{Y}$ and
$Y=\dep{Y}{X}$.  We can write $\ket{\psi}$ in the form
$\ket{\psi}=\sum_{x}\sqrt{P_X(x)}\ket{x}\ket{\varphi_x}$ and get 
$\rho_B = \sum_x P_X(x)\proj{\varphi_x}$. For the leakage of $\ket{\psi}$
we have:
$\Delta_\psi(P_{X,Y})=S(\rho_B)-I(X;Y)=0$. From the Holevo bound
(Theorem~\ref{holevo}) follows that the states $\{\ket{\varphi_x}\}_x$ form an 
orthonormal basis of their span (since $X=\dep{X}{Y}$, they are all 
different) and that $Y$ captures the result of a measurement in this basis, 
which therefore is the computational basis. Since $Y=\dep{Y}{X}$, we get 
that for each $x$, there is a single $y_x\in\mathcal{Y}$ such that $\ket{\varphi_x}=\ket{y_x}$. 
The primitives $P_{\dep{X}{Y},\dep{Y}{X}}$ and $P_{X,Y}$ are therefore trivial.  \qed

\subsection{Tripartite Embeddings and Proof of Theorem~\ref{no-amplif}}
\label{app:tripartite}
It is natural to generalize the scenario involving only two parties to
the setting where the two players also have access to a particular
\emph{trusted third party} who provides them with classical variables
$X',Y'$ sampled according to distribution $P_{X',Y'}$. The state
produced by purifying Alice's and Bob's actions in such a protocol up
to the final measurement yielding $X$ and $Y$ can without loss of
generality be viewed as a pure state shared among Alice, Bob and an
environment $\ket{\psi}_{EABA'B'}= \sum_e
\sqrt{P_E(e)}\ket{e}_E\otimes\ket{\psi^e}_{ABA'B'}$ . 
We define tripartite embeddings of a
primitive $P_{X,Y}$ analogously to the case of embeddings:

\begin{definition}
\label{corr_impl}
A state $\ket{\psi}=\sum_e P_E(e)\ket{e}_E\otimes
\ket{\psi^e}_{ABA'B'}$ is a \emph{tripartite embedding} of $P_{X,Y}$,
if measuring registers $A$ and $B$ in the computational basis yields
$X,Y$ with distribution $P_{X,Y}$ and the ensemble
$\rho_{ABA'B'}\assign \tr_E\proj{\psi}$ satisfies
$S(X;YB')=S(XA';Y)=I(X;Y)$ .
\end{definition}

The generalization of the notion of leakage to tripartite embeddings
is straightforward:
\begin{definition}
Let $\ket{\psi}\in \hil_{E}\otimes\hil_{ABA'B'}$ be a tripartite embedding of $P_{X,Y}$. 
We define the leakage 
of $\rho_{ABA'B'}\assign \tr_E\proj{\psi}$ viewed as an implementation of $P_{X,Y}$ as
$$\Delta_{\rho_{ABA'B'}}(P_{X,Y}):=\max \left\{ S(X;BB')-I(X;Y) \, , \, S(AA';Y)-I(X;Y) \right\}\, .$$
\end{definition}

The leakage of a tripartite embedding is non-negative, for the same reason 
as in the bipartite case however, it is not necessarily symmetric. 

\begin{lemma}
\label{lm:nonleakprivate}
A non-leaking tripartite embedding $\ket{\psi}_{EABA'B'}$ of $P_{X,Y}$
implements $P_{X,Y}$ ideally (which means: equivalently to the ideal
functionality).
\end{lemma}

\begin{proof}
As we can see below, the statement generalizes Theorem~\ref{thm:nonleaktrivial}. 
Here we assume that in $\ket{\psi}_{EABA'B'}$, Alice's and Bob's entire registers are used to compute $X$ and $Y$ 
i.e., there are no additional registers. This 
is without loss of generality because for any $\tilde{Y}$ capturing the result of measuring only a 
part of Bob's register, we get that 
$$S(X;B)\geq I(X;Y)\geq I(X;\tilde{Y})\, .$$ 
Hence, $\ket{\psi}$ being a non-leaking tripartite embedding of $P_{X,\tilde{Y}}$ 
implies that $\ket{\psi}$ is a non-leaking tripartite embedding of $P_{X,Y}$.
Clearly, also $\ket{\psi}$ implementing $P_{X,Y}$ ideally implies 
that $\ket{\psi}$ implements $P_{X,\tilde{Y}}$ ideally. Therefore, 
showing that if $\ket{\psi}$ is a non-leaking tripartite embedding of 
$P_{X,\tilde{Y}}$ then it implements $P_{X,\tilde{Y}}$ ideally is 
equivalent to showing that if $\ket{\psi}$ is a non-leaking tripartite 
embedding of $P_{X,Y}$ then it implements $P_{X,Y}$ ideally, for 
$Y$ capturing the result of measuring the entire register of Bob. An analogous 
argument holds on Alice's side.  
Therefore, the respective additional registers $A'$ and $B'$ 
of Alice and Bob can be taken trivial.
Because $\ket{\psi}_{EAB}$ is $0$--leaking, we have that
$S(X;B)=I(X;Y)$, which by the Holevo bound (Theorem~\ref{holevo})
implies that we can write
$$\tr_{EA}\proj{\psi}=\sum_x P_{X}(x)\tr_E\proj{\varphi^x}\, ,$$
where all $\tr_E\proj{\varphi^x}$ are simultaneously diagonalizable. If the common diagonal basis of these 
states is $\{\ket{z}\}_z$, then the  cq-state shared between Alice, holding her classical output, and Bob is 
$$\rho_{XB}=\sum_{z'} \left(\sum_x \left(P_{X|f_Z(z)=z'}(x)\proj{x}\right)\right)\sigma^{z'}\, ,$$ 
where $f_Z(Z)\assign\dep{Z}{X}$ and  
$$\sigma^{z'}\assign\sum_{z: f_Z(z)=z'} a_z \proj{z}\, .$$ 
This is a purely classical state, implementing the distribution $P_{X,Z}$ securely on Bob's side.
 Any information 
that Bob can learn about the distribution of $X$ is via the distribution of $\dep{Z}{X}$ that he learns by measuring 
his part. Hence, for the honest measurement of Bob captured by  
$Y$, we have that $X\leftrightarrow \dep{Z}{X} \leftrightarrow \dep{Y}{X}$ is a Markov chain. From 
the assumption $S(X;B)=I(X;Y)$ we get:
$$S(X;B)=I(X;Z)=I(X;\dep{Z}{X})=I(X;Y)=I(X;\dep{Y}{X})\, ,$$
yielding $S(X|\dep{Y}{X})=S(X|\dep{Z}{X}).$  
Due to the Markov chain property, 
$$S(X|\dep{Z}{X},\dep{Y}{X})=S(X|\dep{Z}{X})\, ,$$ 
implying that 
$$S(X|\dep{Y}{X},\dep{Z}{X})=S(X|\dep{Y}{X})\, ,$$ 
i.e. $X\leftrightarrow \dep{Y}{X}\leftrightarrow \dep{Z}{X}$ is also a Markov chain. 
Since both $\dep{Z}{X}$ and $\dep{Y}{X}$ are minimum random variables (see Section~\ref{chap:prelim} for the meaning of ``minimum'') $W_Z$, $W_Y$ such that 
$X\leftrightarrow W_Z\leftrightarrow \dep{Z}{X}$ and $X\leftrightarrow W_Y\leftrightarrow \dep{Y}{X}$ are Markov chains, we get that $\dep{Z}{X}=\dep{Y}{X}$.  
Then $\rho_{XB}$ can be written as:
$$\rho_{XB}=\sum_{y'} \sum_x \left(P_{X|f_Y(y)=y'}(x)\proj{x}\right)\rho^{y'}\, ,$$ 
where the support of each of $\rho^{y'}$ only contains $y$-values such that $f_Y(y)=y'.$  
It follows that then, $\rho_{XB}$ privately implements $P_{X,\dep{Y}{X}}$ on Bob's side. Analogously, 
$S(A;Y)=I(X;Y)$ implies that $\rho_{AY}$ privately implements $P_{\dep{Y}{X},Y}$ on Alice's side. 
In such a case, $\tr_{EA}\proj{\psi}=\tr_{EA}\proj{\psi'}$ and $\tr_{EB}\proj{\psi}=\tr_{EB}\proj{\psi'}$ for $\ket{\psi'}_{EAB}$ satisfying
$$
\ket{\psi'}_{EAB}=\sum_{x',y'}\sqrt{P_{\dep{X}{Y},\dep{Y}{X}}(x',y')}\ket{x',y'}_E\ket{\omega^{x',y'}}_{AB}
$$ 
where 
$$\ket{\omega^{x',y'}}_{AB}=\sum_{x,y:f_X(x)=x',f_Y(y)=y'}\alpha^{x,y}\ket{x,y}\, .$$
For $S(X;B)$ we then get that 
\begin{eqnarray*}
S(X;B)&=&I(\dep{X}{Y};\dep{Y}{X})+\sum_{x',y'}P_{\dep{X}{Y},\dep{Y}{X}}(x',y')S(\tr_A\proj{\omega^{x',y'}})\\
&=&I(X;Y)+\sum_{x',y'}P_{\dep{X}{Y},\dep{Y}{X}}(x',y')S(\tr_A\proj{\omega^{x',y'}}).
\end{eqnarray*}
 
Hence, equality $S(X;B)=I(X;Y)$ can hold only if all $\ket{\omega^{x',y'}}$ are product states, implying 
that from each party's point of view,  a non-leaking tripartite embedding has to be equivalent to 
\begin{equation}
\sum_{x',y'}\sqrt{P_{\dep{X}{Y},\dep{Y}{X}}(x',y')}\ket{x',y'}_E\sum_x \sqrt{P_{X|\dep{X}{Y}=x'}(x)}\ket{x}_A\sum_y\sqrt{P_{Y|\dep{Y}{X}=y'}(y)}\ket{y}_{B}.
\label{small_env}
\end{equation}
Clearly, such a tripartite embedding implements $P_{X,Y}$ ideally. Furthermore, in such a case for 
$\ket{\psi}_{EAB}=\sum_e \ket{e}_E\ket{\psi^e}_{AB}$, each $\ket{\psi^e}$ has to be an embedding of 
a trivial primitive. Since the knowledge of $e$ then enables Bob to learn the value of $\dep{X}{Y}$ 
completely, $S(E|B)\geq H(\dep{X}{Y}|Y)$ needs to hold. Analogously we can show that $S(E|A)\geq H(\dep{Y}{X}|X).$
Notice that in the case of a bipartite embedding, this can only happen if the computational basis is the Schmidt basis 
for both Alice and Bob. It follows that the distribution of $P_{\dep{X}{Y},\dep{Y}{X}}$ is then of the form: 
$P_{\dep{X}{Y},\dep{Y}{X}}(x',y'_{x'})=P_{\dep{X}{Y}}(x')$, where for $x'_0\neq x'_1$, $y'_{x'_0}\neq y'_{x'_1}$. 
Primitive $P_{X,Y}$ is then trivial and
the claim of Theorem~\ref{thm:nonleaktrivial} follows. 
\qed

\end{proof}

\begin{proof}[Proof of Theorem~\ref{no-amplif}]
  Consider a quantum protocol equipped with a black box for
  $P_{X,Y}$. Due to~(\ref{small_env}), from the players' perspectives,
  such a protocol is indistinguishable from a protocol where
  $S(E|A)\leq H(\dep{Y}{X}|X)$ and $S(E|B)\leq H(\dep{X}{Y}|Y)$ during
  the entire protocol execution, with the following black-box
  implementation of $P_{X,Y}$:
\begin{equation}
\label{secure_impl}
\ket{\psi}_{EAB}=\sum_{x',y'}\sqrt{P_{\dep{X}{Y},\dep{Y}{X}}(x',y')}\ket{x',y'}_E\ket{x',y'}_{AB}.
\end{equation} 
The bits that each player receives from a black box for $P_{X,Y}$ are only classically correlated 
with the environment  and with the outcome of the other player. It follows that  
at any moment of the protocol's execution, honest-but-curious players can measure their parts 
of the black box output, store their respective classical outcomes, and proceed further without 
being detected. Such a measurement on Alice's side extracts incomplete information about the 
environment which therefore partially collapses. If the measurement takes place at the beginning 
of the computation, where it is not preceded by any non-invertible operation such as another 
measurement, then Alice's uncertainty about the environment at this point is $H(\dep{Y}{X}|X)$. Since the 
environment remains unaffected during the protocol's run, $S(E|A)$ cannot exceed this value at any time later.

WLOG now assume that $H(\dep{Y'}{X'}|X')>H(\dep{Y}{X}|X)$.
There is a tripartite embedding of $P_{X,Y}$ of the form (\ref{secure_impl}), where 
$S(E|A)=H(\dep{Y}{X}|X)$. We have argued that the protocol for $P_{X',Y'}$ built upon 
such a black box is indistinguishable from the same protocol using a different black box 
for $P_{X,Y}$ and furthermore, $S(E|A)\leq H(\dep{Y}{X}|X)$ during the entire run of the protocol. 
However, in the proof of Lemma~\ref{lm:nonleakprivate} we have shown that in any non-leaking tripartite 
embedding of $P_{X',Y'}$, $S(E|A)\geq H(\dep{Y'}{X'}|X')$ must hold. 
Since $H(\dep{Y'}{X'}|X')>H(\dep{Y}{X}|X)$, the protocol must leak information.
\qed

\end{proof}

\section{Leakage of Universal Primitives}
\label{app:universalleakage}

\subsection{Exact calculations}
First, we look at the leakage of the embeddings of Rabin String OT (\srot). 

\begin{theorem} \label{thm:psrot} Any embedding of \psrot\ is at least
 $(1-O(r2^{-r}))$-leaking. For $r=1$ any embedding is at least
 $(h(\frac{1}{4})-\frac{1}{2}) \approx 0.311$-leaking. Furthermore, 
the leakage is the same  for all embeddings of \psrot.
\end{theorem}
\begin{proof}
Let
$$\ket{\psi}=\frac{1}{2^{\frac{r+1}{2}}}\sum_{x\in\{0,1\}^r}e^{i\theta(x,x)}\ket{xx}+\frac{1}{2^{\frac{r+1}{2}}}\left(\sum_{x\in\{0,1\}^r}e^{i\theta(x,\bot)}\ket{x}\right)\ket{\bot}\, ,$$
where $\bot$ denotes an erasure, be a general form of an embedding of \psrot. 

Define $\ket{\varphi} \assign \frac{1}{2^{r/2}}\sum_{x\in\{0,1\}^r}e^{i\theta(x,\bot)}\ket{x}$. If Bob 
guesses the value of Alice's string successfully, Alice gets an ensemble $\rho^0=\frac{1}{2^r}\sum_{x\in\{0,1\}^r}\ketbra{x}{x}$. If an erasure occurs on Bob's side, Alice gets $\rho^1=\ketbra{\varphi}{\varphi}$. We find $S(A)$ by computing the eigenvalues of $\rho_A \assign \frac{1}{2}(\rho^0+\rho^1)$.

Since $\rho^0=\frac{1}{2^r}\id_A$, $\ket{v}$ is an eigenvector of $\rho_A$ if and only if it is an eigenvector of $\rho^1$. If $\ket{v}$ is an eigenvector of $\rho^1$ then either a) $\ket{v}=e^{i\theta}\ket{\varphi}$ or b) $\bracket{v}{\varphi}=0$. If a) is true then
$$\rho_A\ket{v}=\frac{1}{2}(\rho^0\ket{v}+\rho^1\ket{v})=\frac{1}{2}\left(1+\frac{1}{2^r}\right)\ket{v}\, ,$$
whereas in the case b),
$$\rho_A\ket{v}=\frac{1}{2}(\rho^0\ket{v}+\rho^1\ket{v})=\frac{1}{2^{r+1}}\, .$$
The state $\rho_A$ has eigenvalues $\{\frac{1}{2}+\frac{1}{2^{r+1}},\frac{1}{2^{r+1}}\}$, where $\frac{1}{2^{r+1}}$ has multiplicity $2^r-1$. $S(A)$ can then be computed as follows:
\begin{eqnarray*}
S(A)&=&-\left(\frac{1}{2}+\frac{1}{2^{r+1}}\right)\log\left(\frac{1}{2}+\frac{1}{2^{r+1}}\right)+\frac{2^r-1}{2^{r+1}}(r+1)\\
&= &\left(\frac{1}{2}+\frac{1}{2^{r+1}}\right)\left(1-\frac{1}{\ln{2}\cdot 2^{r}}+o\left(\frac{1}{2^r}\right)\right)+\frac{r+1}{2}-\frac{r+1}{2^{r+1}}= \frac{r}{2}+1-O\left(\frac{r}{2^{r}}\right).
\end{eqnarray*}

Since $I(X;Y)=\frac{r}{2}$, for the leakage we get:
$$\Delta_{\psi}(\psrot)=S(A)-I(X;Y)= 1- O\left(\frac{r}{2^{r}}\right)\, .$$
As we can see, the leakage does not depend on the phase-function $\theta$.
\qed

\end{proof}

In the following theorem we minimize the leakage of an embedding of $\pot$.
\begin{theorem}
\label{thm:pot}
Any $\ket{\psi}\in \emb{\pot}$ is at least $\frac{1}{2}$-leaking. 
The leakage is minimized by the canonical embedding.
\end{theorem}
\begin{proof}
Let
$$\ket{\psi}=\frac{1}{2\sqrt{2}}\sum_{x_0,x_1,c\in\{0,1\}}e^{i\theta(x_0x_1,cx_c)}\ket{x_0x_1}\ket{cx_c}$$
be a regular embedding of \pot. Without loss of generality assume that $\theta(00,00)=0$.  Notice that for the local phase-change transforms
\begin{eqnarray*}
U^A&\assign&\ketbra{00}{00}+{\rm exp}(i\theta(01,00))\ketbra{01}{01}+{\rm   exp}(i(\theta(10,10)-\theta(00,10)))\ketbra{10}{10}\\
&+&{\rm exp}(i(\theta(10,10)+\theta(11,01)-\theta(00,10)-\theta(10,01)))\ketbra{11}{11},\\
U^B&\assign&\ketbra{00}{00}+{\rm exp}(i(\theta(00,10)+\theta(10,01)-\theta(10,10)))\ketbra{01}{01}\\
&+&{\rm exp}(i\theta(00,10))\ketbra{10}{10}+{\rm exp}(i(\theta(01,11)-\theta(01,00)))\ketbra{11}{11},
\end{eqnarray*}
we get
$$U^A\otimes U^B\ket{\psi}=\ket{\psi'}=\frac{1}{2}(\ket{0+}\ket{00}+\ket{1+}\ket{01}+\ket{+0}\ket{10}+\frac{\ket{0}+e^{i\omega}\ket{1}}{\sqrt{2}}\ket{1}\ket{11})\, ,$$
where $\omega=\theta(00,10)+\theta(01,00)+\theta(10,01)+\theta(11,11)-\theta(01,01)-\theta(10,10)-\theta(11,01).$

Let $A'$ denote Alice's quantum system for Alice and Bob sharing
$\ket{\psi'}$. Since $S(A)=S(A')$, we can minimize $S(A')$ in
order to minimize $S(A)$.  Assume that Alice and Bob share $\ket{\psi'}$.  For
Bob's selection bit $c=0$, Alice gets an ensemble
$\rho_0=\frac{1}{2}(\ketbra{0+}{0+}+\ketbra{1+}{1+})$, whereas for $c=1$, she
gets
$\rho_1=\frac{1}{2}(\ketbra{+0}{+0}+(\ket{01}+e^{i\omega}\ket{11})(\bra{01}+e^{-i\omega}\bra{11}))$,
where
$\rho_{A'}=\frac{1}{2}(\rho_0+\rho_1)$. By solving the characteristic
equation of $\rho_{A'}$ we get the set of eigenvalues
$\{\frac{1}{4}(1\pm\cos\frac{\omega}{4}),\frac{1}{4}(1\pm
\sin\frac{\omega}{4})\}$.  $S(A')$ can then be expressed as follows:
$$S(A')=1+\frac{h(\frac{1-\cos(\omega/4)}{2})+h(\frac{1-\sin(\omega/4)}{2})}{2}\, .$$
By computing the second derivative of $f(x)=h(\frac{1-\sqrt{x}}{2})$, we get that $f''(x)\leq 0$ in $[0,1]$, implying that $f$ is concave in $[0,1]$.
For $\alpha\in[0,1]$, Jensen's inequality yields $\frac{f(0)+f(1)}{2}\leq f(\alpha)$, and therefore,
$\frac{f(0)+f(1)}{2}\leq \frac{f(\alpha)+f(1-\alpha)}{2}$.
Consequently, the minimum of $h(\frac{1-\cos(\omega/4)}{2})+h(\frac{1-\sin(\omega/4)}{2})=f(\cos^2\frac{\omega}{4})+f(\sin^2\frac{\omega}{4})$ is achieved for $\omega=0$ and in this
case, $S(A')=\frac{3}{2}$.

Finally, we can conclude that the leakage is minimal for the canonical embedding and
$\Delta_{\psi}(P_{X,Y})=S(A)-I(X;Y)=S(A')-I(X;Y)\geq \frac{3}{2}-1=\frac{1}{2}$.
\qed

\end{proof}

There is also a more direct way to
interpret this quantity in the case of the canonical embedding
$\ket{\psi_{\vec{0}}}$ for \pot: If Alice and Bob share a single copy of
$\ket{\psi_{\vec{0}}}$ then there exist POVMs for both of them which
reveal Bob's selection bit to Alice, and the XOR of Alice's bits to
Bob, both with probability $\frac{1}{2}$.
Let $\ket{\Phi^{\pm}}=\frac{1}{\sqrt{2}}(\ket{00}\pm\ket{11})$,
$\ket{\Psi^{\pm}}=\frac{1}{\sqrt{2}}(\ket{01}\pm\ket{10})$ denote the
Bell states, and $\ket{\pm} \assign \frac{1}{\sqrt{2}}(\ket{0}\pm\ket{1})$.
Observe that the canonical embedding $\ket{\psi_{\vec{0}}}$ of \pot\ can be expressed as follows:
\[ 
\ket{\psi_{\vec{0}}} = \frac{1}{2}\ket{\Psi^{-}}\otimes \frac{\ket{\Psi^{-}}-\ket{\Phi^{-}}}{\sqrt{2}} +
           \frac{1}{2}\ket{\Phi^{-}}\otimes \frac{\ket{\Psi^{+}}-\ket{\Phi^{+}}}{\sqrt{2}} +
            \frac{1}{\sqrt{2}} \ket{++}\ket{++}.
\]
In order to get the value $x_0\oplus x_1$ of Alice's bits $x_0$ and $x_1$, 
Bob can use  POVM ${\sf B}=\{{\sf B}_0,{\sf B}_1,{\sf B}_?\}$ where
${\sf B}_0 \assign  
\frac{1}{2}(\ket{\Psi^-}-\ket{\Phi^-})(\bra{\Psi^-}-\bra{\Phi^-})$,
${\sf B}_1  \assign  
\frac{1}{2}(\ket{\Psi^+}-\ket{\Phi^+})(\bra{\Psi^+}-\bra{\Phi^+})$,
and
${\sf B}_{?}  \assign  \proj{++}$. It is easy to verify that Bob gets
outcome ${\sf B}_z$ for $z\in\{0,1\}$ (in which
case $x_0\oplus x_1= z$ with certainty) with probability $\frac{1}{2}$.
Alice's POVM can be defined as ${\sf A}=\{{\sf A}_0,{\sf A}_1,{\sf A}_?\}$ where
${\sf A}_0  \assign  \proj{-+}$,
${\sf A}_1  \assign  \proj{+-}$, and
${\sf A}_{?}  \assign  \id_2-{\sf A}_0-{\sf A}_1$.
By inspection we easily find that 
the probability for Alice to get Bob's selection bit 
is $1-\trace{({\sf A}_?\otimes\id_2)\proj{\psi_{\vec{0}}}} = \frac{1}{2}$.
For any regular embedding of \pot\ we can construct similar POVMs revealing the XOR 
of Alice's bits to Bob and Bob's selection bit to Alice with probability strictly 
more than $\frac{1}{4}$.

\subsection{Lower Bounds}
\begin{theorem}
\label{stringotleakage}
Any embedding $\ket{\psi}$ of \psot\ is  $(1-O(r2^{-r}))$-leaking.
\end{theorem}
\begin{proof}
We use Theorem~\ref{reduction} to show that any (regular) embedding of \psot\ leaks at least as much as some regular embedding of \psrot. Let $(A_0,A_1)$ and $B$ denote Alice's and Bob's respective registers.
Then $\ket{\psi}_{A_0A_1B}\in \emb{\psot}$ can be written in the form:
$$\ket{\psi}=\frac{1}{2^{r/2}}\sum_{x\in\{0,1\}^r} \ket{x}^{A_1}\ket{\psi^x}_{A_0B}\, ,$$ 
where each 
$$\ket{\psi^x}=\frac{1}{2^{(r+1)/2}}\sum_{x'\in\{0,1\}^r}\left(e^{i\theta(x',x,0)}\ket{x'}^{A_0}\ket{0,x'}^B+
e^{i\theta(x',x,1)}\ket{x'}_{A_0}\ket{1,x}^B\right)$$ 
can be viewed as a regular embedding of \psrot. According to Theorem~\ref{reduction} and Theorem~\ref{thm:psrot}, 
we get that 
$$\Delta_{\psot}\geq \Delta_{\psrot}=1-O(r/2^r)\, .$$
\qed
\end{proof}

\begin{theorem}
\label{noisyotleakage2}
If $p<\frac{1}{2}-\frac{1}{2\sqrt{2}}$ then 
$\Delta_{\potn}\geq \frac{\left(1/2-p-\sqrt{p(1-p)}\right)^2}{8\ln 2}.$
\end{theorem}
\begin{proof}
Before starting with the actual proof, we formulate a useful statement, relating two measures of uncertainty of 
a quantum ensemble. 

\begin{theorem}[Average Encoding Theorem~\cite{KNTZ01}]
\label{aver_encoding}
Let $B$ denote a quantum system storing the quantum part of a cq-state $\rho_{XE}= \sum_{x\in\mathcal{X}}P_X(x)\proj{x}\otimes \rho^x_E$. Then 
$$\sum_x P_X(x)\|\rho_E-\rho^x_E\|_1\leq\sqrt{2(\ln 2)S(X;B)}\, .$$
\end{theorem}

Let us start with the proof of Theorem~\ref{noisyotleakage2}. First, we show that for any regular embedding of $P_{X,Y_0Y_1}$ such that 
$Y_0$ and $Y_1$ are independent, 
$$S(A;Y_0Y_1)\leq S(A;Y_0)+S(A;Y_1)\, .$$ 

We can write 
\begin{eqnarray}
S(A;Y_0)+S(A;Y_1)&=&H(Y_0)+H(Y_1)-S(Y_0|A)-S(Y_1|A)\nonumber\\
&=&H(Y_0Y_1)-S(Y_0|A)-S(Y_1|A)\nonumber\\
&\leq& H(Y_0Y_1)-S(Y_0Y_1|A)=S(A;Y_0Y_1)\label{jointentropy}.
\end{eqnarray} 

Let $X,Y_0,Y_1$ be random variables corresponding to Alice's pair of bits, Bob's selection bit, and its value, respectively. For $\potn$ we have that $I(X;Y_0Y_1)=1-h(p)$. $S(A;Y_0Y_1)$ can then be lower-bounded by
$$S(A;Y_0Y_1)\geq S(A;Y_0)+S(A;Y_1)\geq S(A;Y_0)+(1-h(p))\, .$$ 
Hence, for computing the lower bound on $S(A;Y_0Y_1)$, we only need 
to compute the lower bound on $S(A;Y_0)$. 
A state $\ket{\psi}\in \emb{\potn}$ can be written as 
$$\ket{\psi}=\frac{1}{\sqrt{2}}(\ket{\psi_0}^{AB_1}\ket{0}^{B_0}+\ket{\psi_1}^{AB_1}\ket{1}^{B_0})\, .$$ 

Let $\rho^0_A\assign \tr_{B_1}\proj{\psi_0}$ and $\rho^1_A\assign 
\tr_{B_1}\proj{\psi_1}.$ 

By applying Theorem~\ref{aver_encoding} from above, 
we get that 
$$\|\rho^0_A-\rho^1_A\|_1\leq \sqrt{8(\ln 2)S(A;Y_0)}\, ,$$
and therefore,
\begin{equation}
\label{averageencoding}
\frac{\|\rho^0_A-\rho^1_A\|_1^2}{8\ln 2}\leq S(A;Y_0).
\end{equation}
The trace norm of $\rho^0_A-\rho^1_A$ yields an upper bound on the entries of the matrix:
\begin{equation}
\label{norma}
|(\rho^0_A-\rho^1_A)_{ij}|\leq \|\rho^0_A-\rho^1_A\|_1.
\end{equation} 

We can write the state $\ket{\psi}$ in the form:
$$\ket{\psi}=\frac{1}{2}\sum_{y_0,y_1}\ket{\varphi^{y_0,y_1}}_A\ket{y_0,y_1}_{B_0B_1}\, ,$$ 
where 
\begin{eqnarray*}
\ket{\varphi_{0,y}}&=&\sqrt{\frac{1-p}{2}}\sum_{x=0}^1 e^{i\theta(y,x,0,y)}
\ket{y,x}_A\ket{0,y}_{B_0B_1}+\sqrt{\frac{p}{2}}\sum_{x=0}^1 e^{i\theta(y,x,0,1-y)}\ket{y,x}_A\ket{0,1-y}_{B_0B_1}\\
\ket{\varphi_{1,y}}&=&\sqrt{\frac{1-p}{2}}\sum_{x=0}^1 e^{i\theta(x,y,1,y)}
\ket{x,y}_A\ket{1,y}_{B_0B_1}+\sqrt{\frac{p}{2}}\sum_{x=0}^1 e^{i\theta(x,y,1,1-y)}\ket{x,y}_A\ket{1,1-y}_{B_0B_1}.
\end{eqnarray*} 
By evaluating the entries of $(\rho^0_A-\rho^1_A)$ we get a simple lower bound on $|(\rho^0_A-\rho^1_A)_{ij}|$ for $i\neq j\in\{0,\dots,3\}$: 
\begin{equation}
\label{upboundentry}
|(\rho^0_A-\rho^1_A)_{ij}|\geq \frac{1-2p}{4}-\frac{\sqrt{(1-p)p}}{2}
\end{equation}
hence, from~(\ref{norma}) follows that 
$$\|\rho^0_A-\rho^1_A\|_1\geq  \frac{1-2p}{4}-\frac{\sqrt{(1-p)p}}{2}\, ,$$
yielding due to~(\ref{jointentropy}) and (\ref{averageencoding})  that 

$$S(A;Y_0Y_1)\geq 1-h(p)+S(A;Y_0)\geq  1-h(p)+\frac{(1/2-p-\sqrt{(1-p)p})^2}{32\ln 2}\, .$$ 

The lower-bound is non-trivial if $1/2-p-\sqrt{(1-p)p}>0$, 
which is true for $p<\frac{1}{2}-\frac{1}{2\sqrt{2}}$. 
The results yields the following lower-bound on the leakage of \potn: 
$$\Delta_{\potn} \geq \frac{(1/2-p-\sqrt{(1-p)p})^2}{32\ln 2}\, .$$ 
However, this lower-bound is very loose, since for $p=0$ we get 
that 
$$\Delta_{\pot}\geq \frac{1}{128\ln 2}\approx 0.011\, ,$$
which is much weaker than the optimal 
$$\Delta_{\pot}\geq \frac{1}{2}\, .$$

It remains to mention that by using more careful analysis of the phases of $\ket{\varphi_{0.y}}$ and $\ket{\varphi_{1,y}}$, the lower bound on the absolute value of the outside-diagonal 
entries from~(\ref{upboundentry}) can be improved, yielding a non-trivial lower bound on the leakage for $p>0.15$
and eventually, even for any $p<1/4$. It is possible that for the values of $p$ close to $1/4$, we can get a lower bound with a better ratio compared to the real value of the minimum leakage of an embedding of \potn.
\qed

\end{proof}
}{}
\end{document}